\documentclass[12pt]{article}
\pdfoutput=1
\usepackage[backref=page]{hyperref}
\usepackage{xcolor}
\usepackage[]{putex}
\usepackage[utf8]{inputenc}
\usepackage{cite}

\usepackage{amsmath,amsthm,amssymb}
\usepackage{tikz}
\usepackage{tikz-cd}
\usetikzlibrary{decorations.markings}
\usetikzlibrary{calc}
\usepackage{subcaption}
\usepackage{enumerate}
\usepackage{enumitem}

\usepackage{xifthen}

\numberwithin{equation}{section}

\makeatletter
\DeclareRobustCommand*{\bfseries}{%
  \not@math@alphabet\bfseries\mathbf
  \fontseries\bfdefault\selectfont
  \boldmath
}
\makeatother
\hypersetup{
  colorlinks=true,
  linkcolor=blue,
  citecolor=green!50!black,
  urlcolor=cyan!70!black,
  linktoc=all
  }

\input{glyphtounicode}
\def\musepic#1{\vcenter{\hbox{\usebox{#1}}}}

\newsavebox{\figFourCB}
\savebox{\figFourCB}{%
    \begin{tikzpicture}[scale=1]
    \coordinate (y1) at (-7/4,0);
    \coordinate (y2) at (-1,0);
    \coordinate (x3) at (-1,3/4);
    \coordinate (y3) at (0,0);
    \coordinate (x6) at (0,3/4);
    \coordinate (y6) at (3/4,0);
	\draw[thick] (y1)--(y2)--(x3);
	\draw[thick] (y3)--(x6);
	\draw[thick] (y3)--(y6);
	\draw (y1) node[anchor=east] {\scriptsize $\Delta_1$};
	\draw (x3) node[anchor=south] {\scriptsize $\Delta_2$};
	\draw (x6) node[anchor=south] {\scriptsize $\Delta_3$};
	\draw (y6) node[anchor=west] {\scriptsize $\Delta_4$};
	\draw[thick] (y2)--(y3); 
	\draw ($(y2)!0.5!(y3)$) node[anchor=north] {\scriptsize $(\Delta, \ell)$};
\end{tikzpicture}
    }
\newsavebox{\figFiveCB}
\savebox{\figFiveCB}{%
    \begin{tikzpicture}[scale=.75]
    \coordinate (y1) at (-3/2-1/4,0);
    \coordinate (y2) at (-1,0);
    \coordinate (x3) at (-1,1-1/4);
    \coordinate (x4) at (0,1-1/4);
    \coordinate (y3) at (0,0);
    \coordinate (x6) at (1,1-1/4);
    \coordinate (y4) at (1,0);
    \coordinate (y6) at (1+1/2+1/4,0);
	\draw[thick] (y1)--(y2)--(x3);
	\draw[thick] (y3)--(x4);
	\draw[thick] (y4)--(x6);
	\draw[thick] (y4)--(y6);
	\draw (y1) node[anchor=east] {\scriptsize $\Delta_1$};
	\draw (x3) node[anchor=south] {\scriptsize $\Delta_2$};
	\draw (x4) node[anchor=south] {\scriptsize $\Delta_3$};
	\draw (x6) node[anchor=south] {\scriptsize $\Delta_4$};
	\draw (y6) node[anchor=west] {\scriptsize $\Delta_5$};
	\draw[thick] (y2)--(y3); 
	\draw ($(y2)!0.5!(y3)$) node[anchor=north] {\scriptsize $\Delta_{\delta_1}$};
	\draw[thick] (y3)--(y4); 
	\draw ($(y3)!0.5!(y4)$) node[anchor=north] {\scriptsize $\Delta_{\delta_2}$};
\end{tikzpicture}
    }
\newsavebox{\figSixOPECB}
\savebox{\figSixOPECB}{%
    \begin{tikzpicture}[scale=.75]
    \coordinate (y1) at (-3/2,-1/2);
    \coordinate (y2) at (-1,0);
    \coordinate (x3) at (-3/2,1/2);
    \coordinate (y3) at (0,0);
    \coordinate (x4) at (0,1);
    \coordinate (w1) at (-1/2,3/2);
    \coordinate (w2) at (1/2,3/2);
    \coordinate (y4) at (1,0);
    \coordinate (x6) at (3/2,1/2);
    \coordinate (y6) at (3/2,-1/2);
	\draw[thick] (y1)--(y2)--(x3);
	\draw[thick] (w1)--(x4)--(w2);
	\draw[thick] (y4)--(x6);	
	\draw[thick] (y4)--(y6);
	\draw (y1) node[anchor=east] {\scriptsize $\Delta_1$};
	\draw (x3) node[anchor=east] {\scriptsize $\Delta_2$};
	\draw (w1) node[anchor=south] {\scriptsize $\Delta_3$};
	\draw (w2) node[anchor=south] {\scriptsize $\Delta_4$};
	\draw (x6) node[anchor=west] {\scriptsize $\Delta_5$};
	\draw (y6) node[anchor=west] {\scriptsize $\Delta_6$};
    \draw[thick] (y2)--(y3); 
    \draw ($(y2)!0.5!(y3)$) node[anchor=north] {\scriptsize $\Delta_{\delta_1}$};
	\draw[thick] (y3)--(x4); 
	\draw ($(y3)!0.5!(x4)$) node[anchor=west] {\scriptsize $\Delta_{\delta_2}$};
	\draw[thick] (y3)--(y4); 
	\draw ($(y3)!0.5!(y4)$) node[anchor=north] {\scriptsize $\Delta_{\delta_3}$};
\end{tikzpicture}
}
\newsavebox{\figSixCombCB}
\savebox{\figSixCombCB}{%
   \begin{tikzpicture}[scale=.75]
    \coordinate (y1) at (-3/2-1/4,0);
    \coordinate (y2) at (-1,0);
    \coordinate (x3) at (-1,1-1/4);
    \coordinate (x4) at (0,1-1/4);
    \coordinate (x5) at (2,1-1/4);
    \coordinate (y3) at (0,0);
    \coordinate (x6) at (1,1-1/4);
    \coordinate (y4) at (1,0);
    \coordinate (y5) at (2,0);
    \coordinate (y6) at (2+1/2+1/4,0);
	\draw[thick] (y1)--(y2)--(x3);
	\draw[thick] (y3)--(x4);
	\draw[thick] (y4)--(x6);
	\draw[thick] (y5)--(y6);
	\draw[thick] (y5)--(x5);
	\draw (y1) node[anchor=east] {\scriptsize $\Delta_1$};
	\draw (x3) node[anchor=south] {\scriptsize $\Delta_2$};
	\draw (x4) node[anchor=south] {\scriptsize $\Delta_3$};
	\draw (x6) node[anchor=south] {\scriptsize $\Delta_4$};
	\draw (x5) node[anchor=south] {\scriptsize $\Delta_5$};
	\draw (y6) node[anchor=west] {\scriptsize $\Delta_6$};
	\draw[thick] (y2)--(y3); 
	\draw ($(y2)!0.5!(y3)$) node[anchor=north] {\scriptsize $\Delta_{\delta_1}$};
	\draw[thick] (y3)--(y4); 
	\draw ($(y3)!0.5!(y4)$) node[anchor=north] {\scriptsize $\Delta_{\delta_2}$};
	\draw[thick] (y4)--(y5); 
	\draw ($(y4)!0.5!(y5)$) node[anchor=north] {\scriptsize $\Delta_{\delta_3}$};
\end{tikzpicture}
    }
\newsavebox{\figSevenCombCB}
\savebox{\figSevenCombCB}{%
   \begin{tikzpicture}[scale=.75]
    \coordinate (y1) at (-3/2,0);
    \coordinate (y2) at (-1,0);
    \coordinate (x3) at (-1,1-1/4);
    \coordinate (x4) at (0,1-1/4);
    \coordinate (x5) at (2,1-1/4);
    \coordinate (y3) at (0,0);
    \coordinate (x6) at (1,1-1/4);
    \coordinate (y4) at (1,0);
    \coordinate (y5) at (2,0);
    \coordinate (y6) at (2+1/2+1,0);
    \coordinate (z1) at (2+1/2+1/2,0);
    \coordinate (z2) at (2+1/2+1/2,1-1/4);
	\draw[thick] (y1)--(y2)--(x3);
	\draw[thick] (y3)--(x4);
	\draw[thick] (y4)--(x6);
	\draw[thick] (y5)--(y6);
	\draw[thick] (y5)--(x5);
	\draw[thick] (z1)--(z2);
	\draw (y1) node[anchor=east] {\scriptsize $\Delta_1$};
	\draw (x3) node[anchor=south] {\scriptsize $\Delta_2$};
	\draw (x4) node[anchor=south] {\scriptsize $\Delta_3$};
	\draw (x6) node[anchor=south] {\scriptsize $\Delta_4$};
	\draw (x5) node[anchor=south] {\scriptsize $\Delta_5$};
	\draw (z2) node[anchor=south] {\scriptsize $\Delta_6$};
	\draw (y6) node[anchor=west] {\scriptsize $\Delta_7$};
	\draw[thick] (y2)--(y3); 
	\draw ($(y2)!0.5!(y3)$) node[anchor=north] {\scriptsize $\Delta_{\delta_1}$};
	\draw[thick] (y3)--(y4); 
	\draw ($(y3)!0.5!(y4)$) node[anchor=north] {\scriptsize $\Delta_{\delta_2}$};
	\draw[thick] (y4)--(y5); 
	\draw ($(y4)!0.5!(y5)$) node[anchor=north] {\scriptsize $\Delta_{\delta_3}$};
	\draw ($(z1)!0.5!(y5)$) node[anchor=north] {\scriptsize $\Delta_{\delta_4}$};
\end{tikzpicture}
    }
\newsavebox{\figSevenMixedCB}
\savebox{\figSevenMixedCB}{%
   \begin{tikzpicture}[scale=.75]
    \coordinate (y1) at (-3/2,0);
    \coordinate (y2) at (-1,0);
    \coordinate (x3) at (-1,1);
    \coordinate (y3) at (0,0);
    \coordinate (z3) at (0,1/2);
    \coordinate (x4) at (0,1);
    \coordinate (y4) at (1,0);
    \coordinate (z4) at (1,1/2);
    \coordinate (x5) at (1,1);
    \coordinate (y5) at (2,0);
    \coordinate (x6) at (2,1);
    \coordinate (y6) at (5/2,0);
    \coordinate (w3) at (1/2,3/2);
    \coordinate (w4) at (3/2,3/2);
	\draw[thick] (y1)--(y2)--(x3);
	\draw[thick] (y3)--(x4);
	\draw[thick] (y5)--(x6);	
	\draw[thick] (y5)--(y6);
	\draw[thick] (w3)--(x5)--(w4);
	\draw (y1) node[anchor=east] {\scriptsize $\Delta_1$};
	\draw (x3) node[anchor=south] {\scriptsize $\Delta_2$};
	\draw (x4) node[anchor=south] {\scriptsize $\Delta_3$};
	\draw (w3) node[anchor=south] {\scriptsize $\Delta_4$};
	\draw (w4) node[anchor=south] {\scriptsize $\Delta_5$};
	\draw (x6) node[anchor=south] {\scriptsize $\Delta_6$};
	\draw (y6) node[anchor=west] {\scriptsize $\Delta_7$};
	\draw[thick] (y2)--(y3); 
	\draw ($(y2)!0.5!(y3)$) node[anchor=north] {\scriptsize $\Delta_{\delta_1}$};
	\draw[thick] (y3)--(y4); 
	\draw ($(y3)!0.5!(y4)$) node[anchor=north] {\scriptsize $\Delta_{\delta_2}$};
	\draw[thick] (x5)--(y4); 
	\draw ($(x5)!0.5!(y4)$) node[anchor=west] {\scriptsize $\Delta_{\delta_3}$};
	\draw[thick] (y5)--(y4); 
	\draw ($(y5)!0.5!(y4)$) node[anchor=north] {\scriptsize $\Delta_{\delta_4}$};
\end{tikzpicture} 
    }
\newsavebox{\figEightCombCB}
\savebox{\figEightCombCB}{%
   \begin{tikzpicture}[scale=.75]
    \coordinate (y1) at (-3/2,0);
    \coordinate (y2) at (-1,0);
    \coordinate (x3) at (-1,1-1/4);
    \coordinate (x4) at (0,1-1/4);
    \coordinate (x5) at (2,1-1/4);
    \coordinate (y3) at (0,0);
    \coordinate (x6) at (1,1-1/4);
    \coordinate (y4) at (1,0);
    \coordinate (y5) at (2,0);
    \coordinate (y6) at (2+1/2+1+1,0);
    \coordinate (z1) at (2+1/2+1/2+1,0);
    \coordinate (z2) at (2+1/2+1/2+1,1-1/4);
    \coordinate (w1) at (2+1/2+1/2,0);
    \coordinate (w2) at (2+1/2+1/2,1-1/4);
	\draw[thick] (y1)--(y2)--(x3);
	\draw[thick] (y3)--(x4);
	\draw[thick] (y4)--(x6);
	\draw[thick] (y5)--(y6);
	\draw[thick] (y5)--(x5);
	\draw[thick] (z1)--(z2);
	\draw[thick] (w1)--(w2);
	\draw (y1) node[anchor=east] {\scriptsize $\Delta_1$};
	\draw (x3) node[anchor=south] {\scriptsize $\Delta_2$};
	\draw (x4) node[anchor=south] {\scriptsize $\Delta_3$};
	\draw (x6) node[anchor=south] {\scriptsize $\Delta_4$};
	\draw (x5) node[anchor=south] {\scriptsize $\Delta_5$};
	\draw (w2) node[anchor=south] {\scriptsize $\Delta_6$};
	\draw (z2) node[anchor=south] {\scriptsize $\Delta_7$};
	\draw (y6) node[anchor=west] {\scriptsize $\Delta_8$};
	\draw[thick] (y2)--(y3); 
	\draw ($(y2)!0.5!(y3)$) node[anchor=north] {\scriptsize $\Delta_{\delta_1}$};
	\draw[thick] (y3)--(y4); 
	\draw ($(y3)!0.5!(y4)$) node[anchor=north] {\scriptsize $\Delta_{\delta_2}$};
	\draw[thick] (y4)--(y5); 
	\draw ($(y4)!0.5!(y5)$) node[anchor=north] {\scriptsize $\Delta_{\delta_3}$};
	\draw ($(y5)!0.5!(w1)$) node[anchor=north] {\scriptsize $\Delta_{\delta_4}$};
	\draw ($(z1)!0.5!(w1)$) node[anchor=north] {\scriptsize $\Delta_{\delta_5}$};
\end{tikzpicture}
    }
\newsavebox{\figEightOPECB}
\savebox{\figEightOPECB}{%
   \begin{tikzpicture}[scale=.75]
    \coordinate (y1) at (-3/2-3/4,0);
    \coordinate (y2) at (-1-3/4,0);
    \coordinate (x3) at (-1-3/4,1);
    \coordinate (y3) at (-3/4,0);
    \coordinate (z3) at (0,1/2);
    \coordinate (x4) at (-3/4,1);
    \coordinate (y4) at (1,0);
    \coordinate (z4) at (1,1/2);
    \coordinate (x5) at (1,1);
    \coordinate (y5) at (2,0);
    \coordinate (x6) at (2,1);
    \coordinate (y6) at (5/2,0);
    \coordinate (w3) at (1/2,3/2);
    \coordinate (w4) at (3/2,3/2);
    \coordinate (ww3) at (-3/4-1/2,3/2);
    \coordinate (ww4) at (-3/4+1/2,3/2);
	\draw[thick] (y1)--(y2)--(x3);
	\draw[thick] (y3)--(x4);
	\draw[thick] (y5)--(x6);	
	\draw[thick] (y5)--(y6);
	\draw[thick] (w3)--(x5)--(w4);
	\draw[thick] (ww3)--(x4)--(ww4);
	\draw (y1) node[anchor=east] {\scriptsize $\Delta_1$};
	\draw (x3) node[anchor=south] {\scriptsize $\Delta_2$};
	\draw (ww3) node[anchor=south] {\scriptsize $\Delta_3$};
	\draw (ww4) node[anchor=south] {\scriptsize $\Delta_4$};
	\draw (w3) node[anchor=south] {\scriptsize $\Delta_5$};	
	\draw (w4) node[anchor=south] {\scriptsize $\Delta_6$};
	\draw (x6) node[anchor=south] {\scriptsize $\Delta_7$};
	\draw (y6) node[anchor=west] {\scriptsize $\Delta_8$};
	\draw[thick] (y2)--(y3); 
	\draw ($(y2)!0.5!(y3)$) node[anchor=north] {\scriptsize $\Delta_{\delta_1}$};
	\draw[thick] (y3)--(y4); 
	\draw ($(y3)!0.5!(x4)$) node[anchor=west] {\scriptsize $\Delta_{\delta_2}$};
	\draw ($(y3)!0.5!(y4)$) node[anchor=north] {\scriptsize $\Delta_{\delta_3}$};
	\draw[thick] (x5)--(y4); 
	\draw ($(x5)!0.5!(y4)$) node[anchor=west] {\scriptsize $\Delta_{\delta_4}$};
	\draw[thick] (y5)--(y4); 
	\draw ($(y5)!0.5!(y4)$) node[anchor=north] {\scriptsize $\Delta_{\delta_5}$};
\end{tikzpicture} 
    }
\newsavebox{\figEightSymCB}
\savebox{\figEightSymCB}{%
   \begin{tikzpicture}[scale=.75]
    \coordinate (y1) at (-3/2-3/4,0);
    \coordinate (y2) at (-1-3/4,0);
    \coordinate (x3) at (-1-3/4,1);
    \coordinate (y3) at (-3/4+3/4,0);
    \coordinate (z3) at (0,1/2);
    \coordinate (x4) at (-3/4+3/4,1);
    \coordinate (y4) at (1,0);
    \coordinate (z4) at (1,1/2);
    \coordinate (x5) at (1,1);
    \coordinate (y5) at (2,0);
    \coordinate (x6) at (2,1);
    \coordinate (y6) at (5/2,0);
    \coordinate (w3) at (-7/8,0);
    \coordinate (w4) at (-7/8,1);
    \coordinate (ww3) at (-3/4-1/2+3/4,3/2);
    \coordinate (ww4) at (-3/4+1/2+3/4,3/2);
	\draw[thick] (y1)--(y2)--(x3);
	\draw[thick] (y3)--(x4);
	\draw[thick] (y5)--(x6);	
	\draw[thick] (y5)--(y6);
	\draw[thick] (w3)--(w4);
	\draw[thick] (ww3)--(x4)--(ww4);
	\draw (y1) node[anchor=east] {\scriptsize $\Delta_1$};
	\draw (x3) node[anchor=south] {\scriptsize $\Delta_2$};
	\draw (ww3) node[anchor=south] {\scriptsize $\Delta_4$};
	\draw (ww4) node[anchor=south] {\scriptsize $\Delta_5$};
	\draw (w4) node[anchor=south] {\scriptsize $\Delta_3$};	
	\draw (x5) node[anchor=south] {\scriptsize $\Delta_6$};
	\draw (x6) node[anchor=south] {\scriptsize $\Delta_7$};
	\draw (y6) node[anchor=west] {\scriptsize $\Delta_8$};
	\draw[thick] (y2)--(y3); 
	\draw ($(y2)!0.5!(w3)$) node[anchor=north] {\scriptsize $\Delta_{\delta_1}$};
	\draw[thick] (y3)--(y4); 
	\draw ($(y3)!0.5!(x4)$) node[anchor=west] {\scriptsize $\Delta_{\delta_3}$};
	\draw ($(y3)!0.5!(y4)$) node[anchor=north] {\scriptsize $\Delta_{\delta_4}$};
	\draw[thick] (x5)--(y4); 
	\draw ($(w3)!0.5!(y3)$) node[anchor=north] {\scriptsize $\Delta_{\delta_2}$};
	\draw[thick] (y5)--(y4); 
	\draw ($(y5)!0.5!(y4)$) node[anchor=north] {\scriptsize $\Delta_{\delta_5}$};
\end{tikzpicture} 
    }
\newsavebox{\figEightAsymCB}
\savebox{\figEightAsymCB}{%
   \begin{tikzpicture}[scale=.75]
    \coordinate (y1) at (-3/2-3/4,0);
    \coordinate (y2) at (-1-3/4,0);
    \coordinate (x3) at (-1-3/4,1);
    \coordinate (y3) at (-3/4,0);
    \coordinate (z3) at (0,1/2);
    \coordinate (x4) at (-3/4,1);
    \coordinate (y4) at (1,0);
    \coordinate (z4) at (1,1/2);
    \coordinate (x5) at (1,1);
    \coordinate (y5) at (2,0);
    \coordinate (x6) at (2,1);
    \coordinate (y6) at (5/2,0);
    \coordinate (w3) at (1/2,3/2);
    \coordinate (w4) at (3/2,3/2);
    \coordinate (ww3) at (1/8,0);
    \coordinate (ww4) at (1/8,1);
	\draw[thick] (y1)--(y2)--(x3);
	\draw[thick] (y3)--(x4);
	\draw[thick] (y5)--(x6);	
	\draw[thick] (y5)--(y6);
	\draw[thick] (w3)--(x5)--(w4);
	\draw[thick] (ww3)--(ww4);
	\draw (y1) node[anchor=east] {\scriptsize $\Delta_1$};
	\draw (x3) node[anchor=south] {\scriptsize $\Delta_2$};
	\draw (x4) node[anchor=south] {\scriptsize $\Delta_3$};
	\draw (ww4) node[anchor=south] {\scriptsize $\Delta_4$};
	\draw (w3) node[anchor=south] {\scriptsize $\Delta_5$};	
	\draw (w4) node[anchor=south] {\scriptsize $\Delta_6$};
	\draw (x6) node[anchor=south] {\scriptsize $\Delta_7$};
	\draw (y6) node[anchor=west] {\scriptsize $\Delta_8$};
	\draw[thick] (y2)--(y3); 
	\draw ($(y2)!0.5!(y3)$) node[anchor=north] {\scriptsize $\Delta_{\delta_1}$};
	\draw[thick] (y3)--(y4); 
	\draw ($(y3)!0.5!(ww3)$) node[anchor=north] {\scriptsize $\Delta_{\delta_2}$};
	\draw ($(ww3)!0.5!(y4)$) node[anchor=north] {\scriptsize $\Delta_{\delta_3}$};
	\draw[thick] (x5)--(y4); 
	\draw ($(x5)!0.5!(y4)$) node[anchor=west] {\scriptsize $\Delta_{\delta_4}$};
	\draw[thick] (y5)--(y4); 
	\draw ($(y5)!0.5!(y4)$) node[anchor=north] {\scriptsize $\Delta_{\delta_5}$};
\end{tikzpicture} 
    }
\newsavebox{\figFourTmp}

\newcommand{\figCFour}[6]{
\sbox{\figFourTmp}{%
    \begin{tikzpicture}[scale=.75]
    \draw[thick, dashed, fill=black!15!white] (-3+1/4,-2+1+1/4) rectangle (2-1/4,2-1/2);
    \tikzstyle{vint}=[draw,scale=0.4,color=blue,fill=blue,circle]
    \coordinate (y1) at (-7/4,0);
    \coordinate (y2) at (-1,0);
    \coordinate (x3) at (-1,3/4);
    \coordinate (y3) at (0,0);
    \coordinate (x6) at (0,3/4);
    \coordinate (y6) at (3/4,0);
	\draw[thick] (y1)--(y2)--(x3);
	\draw[thick] (y3)--(x6);
	\draw[thick] (y3)--(y6);
	\draw (y1) node[anchor=east] {\scriptsize $#1$};
	\draw (x3) node[anchor=south] {\scriptsize $#2$};
	\draw (x6) node[anchor=south] {\scriptsize $#3$};
	\draw (y6) node[anchor=west] {\scriptsize $#4$};
	\ifthenelse{\isempty{#6}}{\draw[thick] (y2)--(y3); \draw ($(y2)!0.5!(y3)$) node[anchor=north] {\scriptsize $#5$};}{	\draw[thick, densely dashed, red] (y2)--(y3); \draw[red] ($(y2)!0.5!(y3)$) node[anchor=north] {\scriptsize $#5$};};
	\draw (y2) node[vint] {};
	\draw (y3) node[vint] {};
\end{tikzpicture}
    }
\musepic{\figFourTmp}
}
\newsavebox{\figFiveTmp}

\newcommand{\figCFive}[9]{
\sbox{\figFiveTmp}{%
    \begin{tikzpicture}[scale=.75]
    \draw[thick, dashed, fill=black!15!white] (-3+1/4,-2+1+1/4) rectangle (3-1/4,2-1/2);
    \tikzstyle{vint}=[draw,scale=0.4,color=blue,fill=blue,circle]
    \coordinate (y1) at (-3/2-1/4,0);
    \coordinate (y2) at (-1,0);
    \coordinate (x3) at (-1,1-1/4);
    \coordinate (x4) at (0,1-1/4);
    \coordinate (y3) at (0,0);
    \coordinate (x6) at (1,1-1/4);
    \coordinate (y4) at (1,0);
    \coordinate (y6) at (1+1/2+1/4,0);
	\draw[thick] (y1)--(y2)--(x3);
	\draw[thick] (y3)--(x4);
	\draw[thick] (y4)--(x6);
	\draw[thick] (y4)--(y6);
	\draw (y1) node[anchor=east] {\scriptsize $#1$};
	\draw (x3) node[anchor=south] {\scriptsize $#2$};
	\draw (x4) node[anchor=south] {\scriptsize $#3$};
	\draw (x6) node[anchor=south] {\scriptsize $#4$};
	\draw (y6) node[anchor=west] {\scriptsize $#5$};
	\ifthenelse{\isempty{#7}}{\draw[thick] (y2)--(y3); \draw ($(y2)!0.5!(y3)$) node[anchor=north] {\scriptsize $#6$};}{	\draw[thick, densely dashed, red] (y2)--(y3); \draw[red] ($(y2)!0.5!(y3)$) node[anchor=north] {\scriptsize $#6$};};
	\ifthenelse{\isempty{#9}}{\draw[thick] (y3)--(y4); \draw ($(y3)!0.5!(y4)$) node[anchor=north] {\scriptsize $#8$};}{	\draw[thick, densely dashed, red] (y3)--(y4); \draw[red] ($(y3)!0.5!(y4)$) node[anchor=north] {\scriptsize $#8$};};
	\draw (y2) node[vint] {};
	\draw (y3) node[vint] {};
	\draw (y4) node[vint] {};
\end{tikzpicture}
    }
\musepic{\figFiveTmp}
}
\newsavebox{\figSixCombTmp}

\newcommand{\figCSixComb}[6]{
 \def\tempa{#1}%
 \def\tempb{#2}%
 \def\tempc{#3}%
 \def\tempd{#4}%
 \def\tempe{#5}%
 \def\tempf{#6}%
 \figCSixCombCtd
 }
 
\newcommand{\figCSixCombCtd}[6]{
\sbox{\figSixCombTmp}{%
    \begin{tikzpicture}[scale=.75]
    \draw[thick, dashed, fill=black!15!white] (-3+1/4,-2+1+1/4) rectangle (4-1/4,2-1/4-1/4);
    \tikzstyle{vint}=[draw,scale=0.4,color=blue,fill=blue,circle]
    \coordinate (y1) at (-3/2-1/4,0);
    \coordinate (y2) at (-1,0);
    \coordinate (x3) at (-1,1-1/4);
    \coordinate (x4) at (0,1-1/4);
    \coordinate (x5) at (2,1-1/4);
    \coordinate (y3) at (0,0);
    \coordinate (x6) at (1,1-1/4);
    \coordinate (y4) at (1,0);
    \coordinate (y5) at (2,0);
    \coordinate (y6) at (2+1/2+1/4,0);
	\draw[thick] (y1)--(y2)--(x3);
	\draw[thick] (y3)--(x4);
	\draw[thick] (y4)--(x6);
	\draw[thick] (y5)--(y6);
	\draw[thick] (y5)--(x5);
	\draw (y1) node[anchor=east] {\scriptsize $\tempa$};
	\draw (x3) node[anchor=south] {\scriptsize $\tempb$};
	\draw (x4) node[anchor=south] {\scriptsize $\tempc$};
	\draw (x6) node[anchor=south] {\scriptsize $\tempd$};
	\draw (x5) node[anchor=south] {\scriptsize $\tempe$};
	\draw (y6) node[anchor=west] {\scriptsize $\tempf$};
	\ifthenelse{\isempty{#2}}{\draw[thick] (y2)--(y3); \draw ($(y2)!0.5!(y3)$) node[anchor=north] {\scriptsize $#1$};}{	\draw[thick, densely dashed, red] (y2)--(y3); \draw[red] ($(y2)!0.5!(y3)$) node[anchor=north] {\scriptsize $#1$};};
	\ifthenelse{\isempty{#4}}{\draw[thick] (y3)--(y4); \draw ($(y3)!0.5!(y4)$) node[anchor=north] {\scriptsize $#3$};}{	\draw[thick, densely dashed, red] (y3)--(y4); \draw[red] ($(y3)!0.5!(y4)$) node[anchor=north] {\scriptsize $#3$};};
	\ifthenelse{\isempty{#6}}{\draw[thick] (y4)--(y5); \draw ($(y4)!0.5!(y5)$) node[anchor=north] {\scriptsize $#5$};}{	\draw[thick, densely dashed, red] (y4)--(y5); \draw[red] ($(y4)!0.5!(y5)$) node[anchor=north] {\scriptsize $#5$};};
	\draw (y2) node[vint] {};
	\draw (y3) node[vint] {};
	\draw (y4) node[vint] {};
	\draw (y5) node[vint] {};
\end{tikzpicture}
    }
\musepic{\figSixCombTmp}
}
\newsavebox{\figSixOPETmp}

\newcommand{\figCSixOPE}[6]{
 \def\tempa{#1}%
 \def\tempb{#2}%
 \def\tempc{#3}%
 \def\tempd{#4}%
 \def\tempe{#5}%
 \def\tempf{#6}%
 \figCSixOPECtd
 }
 
\newcommand{\figCSixOPECtd}[6]{
\sbox{\figSixOPETmp}{%
    \begin{tikzpicture}[scale=.75]
    \draw[thick, dashed, fill=black!15!white] (-3+1/2,-2+1) rectangle (3-1/2,3-5/8);
    \tikzstyle{vint}=[draw,scale=0.4,color=blue,fill=blue,circle]
    \coordinate (y1) at (-3/2,-1/2);
    \coordinate (y2) at (-1,0);
    \coordinate (x3) at (-3/2,1/2);
    \coordinate (y3) at (0,0);
    \coordinate (x4) at (0,1);
    \coordinate (w1) at (-1/2,3/2);
    \coordinate (w2) at (1/2,3/2);
    \coordinate (y4) at (1,0);
    \coordinate (x6) at (3/2,1/2);
    \coordinate (y6) at (3/2,-1/2);
	\draw[thick] (y1)--(y2)--(x3);
	\draw[thick] (w1)--(x4)--(w2);
	\draw[thick] (y4)--(x6);	
	\draw[thick] (y4)--(y6);
	\draw (y1) node[anchor=east] {\scriptsize $\tempa$};
	\draw (x3) node[anchor=east] {\scriptsize $\tempb$};
	\draw (w1) node[anchor=south] {\scriptsize $\tempc$};
	\draw (w2) node[anchor=south] {\scriptsize $\tempd$};
	\draw (x6) node[anchor=west] {\scriptsize $\tempe$};
	\draw (y6) node[anchor=west] {\scriptsize $\tempf$};
	\ifthenelse{\isempty{#2}}{\draw[thick] (y2)--(y3); \draw ($(y2)!0.5!(y3)$) node[anchor=north] {\scriptsize $#1$};}{	\draw[thick, densely dashed, red] (y2)--(y3); \draw[red] ($(y2)!0.5!(y3)$) node[anchor=north] {\scriptsize $#1$};};
	\ifthenelse{\isempty{#4}}{\draw[thick] (y3)--(x4); \draw ($(y3)!0.5!(x4)$) node[anchor=west] {\scriptsize $#3$};}{	\draw[thick, densely dashed, red] (y3)--(x4); \draw[red] ($(y3)!0.5!(x4)$) node[anchor=west] {\scriptsize $#3$};};
	\ifthenelse{\isempty{#6}}{\draw[thick] (y3)--(y4); \draw ($(y3)!0.5!(y4)$) node[anchor=north] {\scriptsize $#5$};}{	\draw[thick, densely dashed, red] (y3)--(y4); \draw[red] ($(y3)!0.5!(y4)$) node[anchor=north] {\scriptsize $#5$};};
	\draw (y2) node[vint] {};
	\draw (y3) node[vint] {};
	\draw (y4) node[vint] {};
	\draw (x4) node[vint] {};
\end{tikzpicture}
   }
\musepic{\figSixOPETmp}
}
\newsavebox{\figSevenMixTmp}

\newcommand{\figCSevenMix}[7]{
 \def\tempa{#1}%
 \def\tempb{#2}%
 \def\tempc{#3}%
 \def\tempd{#4}%
 \def\tempe{#5}%
 \def\tempf{#6}%
 \def\tempg{#7}%
 \figCSevenMixCtd
 }
 
\newcommand{\figCSevenMixCtd}[8]{
\sbox{\figSevenMixTmp}{%
    \begin{tikzpicture}[scale=.7]
    \draw[thick, dashed, fill=black!15!white] (-3+1/2,-2+1+1/4) rectangle (4-1/2,3-6/8);
    \tikzstyle{vint}=[draw,scale=0.4,color=blue,fill=blue,circle]
    \coordinate (y1) at (-3/2,0);
    \coordinate (y2) at (-1,0);
    \coordinate (x3) at (-1,1);
    \coordinate (y3) at (0,0);
    \coordinate (z3) at (0,1/2);
    \coordinate (x4) at (0,1);
    \coordinate (y4) at (1,0);
    \coordinate (z4) at (1,1/2);
    \coordinate (x5) at (1,1);
    \coordinate (y5) at (2,0);
    \coordinate (x6) at (2,1);
    \coordinate (y6) at (5/2,0);
    \coordinate (w3) at (1/2,3/2);
    \coordinate (w4) at (3/2,3/2);
	\draw[thick] (y1)--(y2)--(x3);
	\draw[thick] (y3)--(x4);
	\draw[thick] (y5)--(x6);	
	\draw[thick] (y5)--(y6);
	\draw[thick] (w3)--(x5)--(w4);
	\draw (y1) node[anchor=east] {\scriptsize $\tempa$};
	\draw (x3) node[anchor=south] {\scriptsize $\tempb$};
	\draw (x4) node[anchor=south] {\scriptsize $\tempc$};
	\draw (w3) node[anchor=south] {\scriptsize $\tempd$};
	\draw (w4) node[anchor=south] {\scriptsize $\tempe$};
	\draw (x6) node[anchor=south] {\scriptsize $\tempf$};
	\draw (y6) node[anchor=west] {\scriptsize $\tempg$};
	\ifthenelse{\isempty{#2}}{\draw[thick] (y2)--(y3); \draw ($(y2)!0.5!(y3)$) node[anchor=north] {\scriptsize $#1$};}{	\draw[thick, densely dashed, red] (y2)--(y3); \draw[red] ($(y2)!0.5!(y3)$) node[anchor=north] {\scriptsize $#1$};};
	\ifthenelse{\isempty{#4}}{\draw[thick] (y3)--(y4); \draw ($(y3)!0.5!(y4)$) node[anchor=north] {\scriptsize $#3$};}{	\draw[thick, densely dashed, red] (y3)--(y4); \draw[red] ($(y3)!0.5!(y4)$) node[anchor=north] {\scriptsize $#3$};};
	\ifthenelse{\isempty{#6}}{\draw[thick] (x5)--(y4); \draw ($(x5)!0.5!(y4)$) node[anchor=west] {\scriptsize $#5$};}{	\draw[thick, densely dashed, red] (x5)--(y4); \draw[red] ($(x5)!0.5!(y4)$) node[anchor=west] {\scriptsize $#5$};};
	\ifthenelse{\isempty{#8}}{\draw[thick] (y5)--(y4); \draw ($(y5)!0.5!(y4)$) node[anchor=north] {\scriptsize $#7$};}{	\draw[thick, densely dashed, red] (y5)--(y4); \draw[red] ($(y5)!0.5!(y4)$) node[anchor=north] {\scriptsize $#7$};};
	\draw (y2) node[vint] {};
	\draw (y3) node[vint] {};
	\draw (y4) node[vint] {};
	\draw (y5) node[vint] {};
    \draw (x5) node[vint] {};
\end{tikzpicture} 
  }
\musepic{\figSevenMixTmp}
}
\newsavebox{\figEdge}

\newcommand{\DrawEdge}[2]{
\savebox{\figEdge}{%
\begin{tikzpicture}[scale=.6]
    \draw[thick, dashed, fill=black!15!white] (0-1/4,-2+1) rectangle (2+1/4,1);
    \tikzstyle{vint}=[draw,scale=0.4,color=blue,fill=blue,circle]
    \coordinate (y1) at (0,0);
    \coordinate (y2) at (2,0);
	\ifthenelse{\isempty{#2}}{\draw[thick] (y1)--(y2);	\draw ($(y1)!0.5!(y2)$) node[anchor=north] {\scriptsize $#1$};}{\draw[thick, dashed, red] (y1)--(y2);	\draw[red] ($(y1)!0.5!(y2)$) node[anchor=north] {\scriptsize $#1$};}
    \draw (y1) node[vint] {};
    \draw (y2) node[vint] {};
\end{tikzpicture}
    }
\musepic{\figEdge}
}
\newsavebox{\figVertex}

\newcommand{\DrawVertex}[6]{
\sbox{\figVertex}{%
    \begin{tikzpicture}[scale=.6]
    \draw[thick, dashed, fill=black!15!white] (0-1/8,-2+1+3/8) rectangle (2+1/8,3-3/8);
    \tikzstyle{vint}=[draw,scale=0.4,color=blue,fill=blue,circle]
    \coordinate (y4) at (1,1/4);
    \coordinate (x5) at (1,1);
    \coordinate (w3) at (1/2,7/4);
    \coordinate (w4) at (3/2,7/4);
	\ifthenelse{\isempty{#6}}{\draw[thick] (y4)--(x5);}{\draw[thick,densely dashed,red] (y4)--(x5);}
	\ifthenelse{\isempty{#2}}{\draw[thick] (w3)--(x5);}{\draw[thick,densely dashed,red] (w3)--(x5);}
	\ifthenelse{\isempty{#4}}{\draw[thick] (w4)--(x5);}{\draw[thick,densely dashed,red] (w4)--(x5);}
	\ifthenelse{\isempty{#6}}{\draw (y4) node[anchor=north] {\scriptsize $#5$};}{\draw[red] (y4) node[anchor=north] {\scriptsize $#5$};}
	\ifthenelse{\isempty{#2}}{\draw (w3) node[anchor=south] {\scriptsize $#1$};}{\draw[red] (w3) node[anchor=south] {\scriptsize $#1$};}	
	\ifthenelse{\isempty{#4}}{\draw (w4) node[anchor=south] {\scriptsize $#3$};}{\draw[red] (w4) node[anchor=south] {\scriptsize $#3$};}
    \draw (x5) node[vint] {};
\end{tikzpicture}
    }
\musepic{\figVertex}
}
\newsavebox{\figNTmp}

\newsavebox{\figNptUnitOne}
\savebox{\figNptUnitOne}{%
\begin{tikzpicture}[scale=1]
    \tikzstyle{blob}=[draw,scale=3,color=black,fill=lightgray,circle]
    \coordinate (x1) at (240:2);
    \coordinate (x2) at (200:2);
    \coordinate (x4) at (-20:2);
    \coordinate (x5) at (-60:2);
    \coordinate (y1) at (130:3);
    \coordinate (y2) at (110:3.5);
    \coordinate (y3) at (70:3.5);
    \coordinate (y4) at (50:3);
    \coordinate (x3) at (110:2.5);
    \coordinate (x6) at (70:2.5);
    \coordinate (y) at (220:1.2);
    \coordinate (yp) at (-40:1.2);
    \coordinate (ypp) at (90:1.5);
    \coordinate (z) at (90:0.2);
    \coordinate (c) at (90:0.85);
    \coordinate (b) at (135:1.2);
	\draw[thick] (x1)--(y)--(x2);
	\draw[thick] (x4)--(yp)--(x5);
	\draw[thick] (y1)--(x3)--(y2);
	\draw[thick] (y3)--(x6)--(y4);
	\draw[thick] (y)--(z)--(yp);
	\draw[thick] (x6)--(ypp)--(x3);
	\draw[thick] (ypp)--(z);
	\draw[thick] (c)--(b);
	\draw (b) node[anchor=east] {\footnotesize $\Delta_1$};
	\draw  ($(ypp)!0.5!(c)$) node[anchor=west] {\footnotesize $\Delta_{2}$};
	\draw  ($(z)!0.5!(c)$) node[anchor=west] {\footnotesize $\Delta_{3}$};
	\draw[very thick, line cap=round, dash pattern=on 0 off 12] (x1)--(x2);
	\draw[very thick, line cap=round, dash pattern=on 0 off 12] (x4)--(x5);
	\draw[very thick, line cap=round, dash pattern=on 0 off 10] (y1)--(y2);
	\draw[very thick, line cap=round, dash pattern=on 0 off 10] (y3)--(y4);
	\draw (y) node[blob] {};
	\draw (yp) node[blob] {};
	\draw (x3) node[blob] {};
	\draw (x6) node[blob] {};
\end{tikzpicture}}
\newsavebox{\figNptUnitTwo}
\savebox{\figNptUnitTwo}{%
\begin{tikzpicture}[scale=1]
    \tikzstyle{blob}=[draw,scale=3,color=black,fill=lightgray,circle]
    \coordinate (x1) at (240:2);
    \coordinate (x2) at (200:2);
    \coordinate (x4) at (-20:2);
    \coordinate (x5) at (-60:2);
    \coordinate (y1) at (130:3);
    \coordinate (y2) at (110:3.5);
    \coordinate (y3) at (70:3.5);
    \coordinate (y4) at (50:3);
    \coordinate (x3) at (110:2.5);
    \coordinate (x6) at (70:2.5);
    \coordinate (y) at (220:1.2);
    \coordinate (yp) at (-40:1.2);
    \coordinate (ypp) at (90:1.5);
    \coordinate (z) at (90:0.2);
    \coordinate (c) at (90:0.85);
    \coordinate (b) at (135:1.2);
	\draw[thick] (x1)--(y)--(x2);
	\draw[thick] (x4)--(yp)--(x5);
	\draw[thick] (y1)--(x3)--(y2);
	\draw[thick] (y3)--(x6)--(y4);
	\draw[thick] (y)--(z)--(yp);
	\draw[thick] (x6)--(ypp)--(x3);
	\draw[thick] (ypp)--(z);
	\draw  ($(ypp)!0.5!(z)$) node[anchor=west] {\footnotesize $\Delta_{3}$};
	\draw[very thick, line cap=round, dash pattern=on 0 off 12] (x1)--(x2);
	\draw[very thick, line cap=round, dash pattern=on 0 off 12] (x4)--(x5);
	\draw[very thick, line cap=round, dash pattern=on 0 off 10] (y1)--(y2);
	\draw[very thick, line cap=round, dash pattern=on 0 off 10] (y3)--(y4);
	\draw (y) node[blob] {};
	\draw (yp) node[blob] {};
	\draw (x3) node[blob] {};
	\draw (x6) node[blob] {};
\end{tikzpicture}}
\newsavebox{\figNptUnitThree}
\savebox{\figNptUnitThree}{%
\begin{tikzpicture}[scale=1]
    \tikzstyle{blob}=[draw,scale=3,color=black,fill=lightgray,circle]
    \coordinate (y3) at (70:3.5);
    \coordinate (y4) at (50:3);
    \coordinate (x3) at (110:2.5);
    \coordinate (x6) at (70:2.5);
    \coordinate (y) at (220:1.2);
    \coordinate (yp) at (-40:1.2);
    \coordinate (ypp) at (90:1.5);
    \coordinate (z) at (90:0.2);
    \coordinate (c) at (90:0.85);
	\draw[thick] (y)--(z)--(yp);
	\draw[thick] (x6)--(ypp)--(x3);
	\draw[thick] (ypp)--(z);
	\draw (x6) node[anchor=south] {\footnotesize $\Delta_2$};
	\draw (x3) node[anchor=south] {\footnotesize $\Delta_1$};
	\draw  ($(ypp)!0.5!(c)$) node[anchor=west] {\footnotesize $\Delta_{3}$};
	\draw[very thick, line cap=round, dash pattern=on 0 off 15] (yp)--(y);
	\draw (z) node[blob] {};
\end{tikzpicture}}
\newsavebox{\figNptUnitFour}
\savebox{\figNptUnitFour}{%
\begin{tikzpicture}[scale=1]
    \tikzstyle{blob}=[draw,scale=3,color=black,fill=lightgray,circle]
    \coordinate (y3) at (70:3.5);
    \coordinate (y4) at (50:3);
    \coordinate (y) at (220:1.2);
    \coordinate (yp) at (-40:1.2);
    \coordinate (ypp) at (90:1.5);
    \coordinate (z) at (90:0.2);
    \coordinate (c) at (90:0.85);
	\draw[thick] (y)--(z)--(yp);
	\draw[thick] (ypp)--(z);
	\draw  (ypp) node[anchor=south] {\footnotesize $\Delta_{3}$};
	\draw[very thick, line cap=round, dash pattern=on 0 off 15] (yp)--(y);
	\draw (z) node[blob] {};
\end{tikzpicture}}

\begin{document}

\title{Dimensional reduction of higher-point conformal blocks}
\authors{Sarah Hoback$^1$\footnote{\tt sarahhoback98@gmail.com} \& Sarthak Parikh$^2$\footnote{\tt sparikh@caltech.edu}}
\institution{PC}{$^1$Department of Physics and Astronomy, Pomona College, Claremont, CA 91711, USA}
\institution{Caltech}{$^2$Division of Physics, Mathematics and Astronomy, California Institute of Technology,\cr\hskip0.06in Pasadena, CA 91125, USA}

\abstract{ 
Recently, with the help of Parisi-Sourlas supersymmetry an intriguing relation was found expressing the four-point scalar conformal block of a $(d-2)$-dimensional CFT in terms of a five-term linear combination of blocks of a $d$-dimensional CFT, with constant coefficients. We extend this dimensional reduction relation to all higher-point scalar conformal blocks of arbitrary topology restricted to scalar exchanges. We show that the constant coefficients appearing in the finite term higher-point dimensional reduction obey an interesting factorization property allowing them to be determined in terms of certain graphical Feynman-like rules and the associated finite set of vertex and edge factors. Notably, these rules can be fully determined by considering the explicit power-series representation of just three particular conformal blocks: the four-point block, the five-point block and the six-point block of the so-called OPE/snowflake topology. In principle, this method can be applied to obtain the arbitrary-point dimensional reduction of conformal blocks with spinning exchanges as well. We also show how to systematically extend the dimensional reduction relation of conformal partial waves to higher-points.
}

\date{September 2020}

\maketitle

\setcounter{tocdepth}{2}

{\hypersetup{linkcolor=black}
\tableofcontents
}


\section{Introduction}
\label{INTRO}

Conformal field theories (CFTs) are important to study as they provide a useful framework for describing a variety of physical phenomena in nature, and offer an insightful window into quantum gravity via the AdS/CFT correspondence. By charting out fixed points of renormalization group flows they also serve as natural guideposts in the space of quantum field theories.
A revival of the bootstrap program~\cite{Ferrara:1973yt,Polyakov:1974gs} has led to  impressive recent progress and powerful new constraints on the space of CFTs in spacetime dimension $d>2$~\cite{Rattazzi:2008pe} (see e.g.\ the review~\cite{Poland:2018epd} and the references therein). 

An important ingredient in the bootstrap program is the knowledge of conformal blocks, which is crucial for setting up the crossing equations.
Conformal blocks are central objects in CFTs as they form a basis for local observables, and via the AdS/CFT correspondence also provide a position space basis for Witten diagrams.
Early foundational work~\cite{Ferrara:1971vh,Ferrara:1972xe,Ferrara:1973vz,Ferrara:1974ny,Polyakov:1974gs} and the seminal works of Dolan and Osborn~\cite{Dolan:2000ut,Dolan:2003hv} established the theory of conformal blocks.
In this paper will focus on global conformal blocks, which are associated with the global conformal group $SO(d+1,1)$. 
Global conformal blocks are eigenfunctions of the conformal Casimir operator of the global conformal group, satisfying appropriate boundary conditions~\cite{Dolan:2000ut,Dolan:2003hv}. 
They also satisfy several interesting mathematical properties~\cite{Dolan:2011dv}, for example various recursion relations (see e.g.\ \cite{Dolan:2003hv,Penedones:2015aga,Kravchuk:2017dzd}), integrability properties (see e.g.\ \cite{Isachenkov:2016gim}), and a geometric interpretation (see e.g.\ \cite{Hijano:2015zsa}).
Closed-form expressions in terms of hypergeometric functions are also available when $d$ is even~\cite{Dolan:2003hv}.
However, until recently, much of the focus has been restricted to four-point blocks which provide a basis for four-point conformal correlators. 

It is useful to go beyond four-point blocks and study higher-point conformal blocks as they are one of the central components of a potential higher-point conformal bootstrap program~\cite{Rosenhaus:2018zqn}, and they provide a canonical direct-channel basis in position-space for all higher-point tree-level Witten diagrams, which are useful for investigating higher-loop effects in AdS/CFT~\cite{Meltzer:2019nbs}.
Over the past two years a spate of new results have appeared in the literature related to higher-point blocks in $d$ dimensions. 
The techniques often used include the shadow formalism~\cite{Rosenhaus:2018zqn}, embedding space OPEs~\cite{Fortin:2016lmf,Fortin:2019fvx,Fortin:2019dnq,Fortin:2019zkm,Fortin:2020ncr,Fortin:2020yjz,Fortin:2020bfq,Fortin:2020zxw},\footnote{See also~\cite{Goncalves:2019znr}.} geodesic Witten diagrams~\cite{Parikh:2019ygo,Jepsen:2019svc,Parikh:2019dvm}
and Mellin space methods~\cite{Hoback:2020pgj}.
Most of the explicit results obtained in literature have been for scalar conformal blocks with scalar exchanges.
The program to determine all higher-point scalar blocks recently culminated in a simple graphical Feynman rules-like prescription for directly writing down any $n$-point block in an arbitrary topology~\cite{Fortin:2020bfq,Hoback:2020pgj,InPrep}.

Still, in contrast to the four-point blocks, only power series expansions in powers of cross-ratios are available and no closed-form expressions for higher-point blocks are known in $d>2$.
Furthermore no explicit results are available for $d$ dimensional conformal blocks involving spinning exchanges.\footnote{Approaches to generalizing the Feynman-rules to include spins may potentially involve the use of weight-shifting operators~\cite{Karateev:2017jgd}, recursive techniques or the embedding space formalism~\cite{Fortin:2019dnq,Fortin:2020ncr}. Recently all higher-point spinning blocks in 2 dimensions were obtained in ref.~\cite{Fortin:2020zxw}.}
One of the objectives of the present work is to step towards filling these gaps.

Despite being seemingly more complicated objects fixed entirely by conformal symmetry, higher-point blocks share several features and properties with their simpler four-point cousins. 
For one, just like the four-point block they admit a power series description dictated by a Feynman-like prescription, which can essentially be read off simply from its unique unrooted binary tree graphical representation~\cite{Hoback:2020pgj}.
For another, they satisfy dimensional reduction relations that are very similar to those obeyed by four-point blocks. 
Just as in the four-point block case, such relations may be hinting at the existence of as yet unknown closed-form expressions for higher-point blocks at higher $d$, e.g.\ for all even $d$. 
While it is not known whether scalar blocks with spinning exchanges also admit a Feynman-like prescription or satisfy dimensional reduction relations, it is strongly suspected that they do, in which case finding such relations may make it possible to explicitly construct all possible conformal blocks in higher $d$ (with or without spinning exchanges) in terms of a small number of basic building blocks.

An indication that such dimensional relations exist at higher-points can be deduced from recent work on Parisi-Sourlas SCFT~\cite{Parisi:1979ka, Kaviraj:2019tbg}.
The SCFT is defined on the superspace $\mathbb{R}^{d|2}$ and transforms under the superconformal group $OSp(d+1,1|2)$. 
Restricting to a $\mathbb{R}^{d-2}$ subspace breaks the superconformal symmetry to $SO(d-1,1) \times OSp(2|2)$ and restricting further to the $OSp(2|2)$-singlet sector gives rise to a {\it local} CFT$_{d-2}$.
It can then be shown that the supermultiplets of the original SCFT$_d$ are in one-to-one correspondence with conformal multiplets of the dimensionally reduced CFT$_{d-2}$.\footnote{Up to certain subtleties; see~ref.\ \cite{Kaviraj:2019tbg} for details.}
Moreover, using the Casimir equation in super-embedding space, it is possible to establish an exact equivalence between the corresponding four-point superconformal block of the SCFT and the the four-point conformal block of the CFT$_{d-2}$.
Since the superconformal block can be written in components in terms of ordinary (non-supersymmetric) $d$ dimensional conformal blocks, this immediately leads to a (finite term) relation between ordinary conformal blocks in $d$ dimensions and $(d-2)$ dimensions:
\eqn{Rych4Ell}{
g_{\Delta,\ell}^{(d-2)} = g_{\Delta,\ell}^{(d)} + c_{2,0}\: g_{\Delta + 2,\ell}^{(d)} + c_{1,-1}\: g_{\Delta+1,\ell-1}^{(d)} + c_{0,-2}\: g_{\Delta,\ell-2}^{(d)} + c_{2,-2}\: g_{\Delta+2,\ell-2}^{(d)} \,,
}
where $c_{i,j}$s are known constants that depend on the external dimensions, the dimension and spin of the exchanged operator and spacetime dimension $d$.
From the representation theory point of view, the terms on the right hand side above correspond to various Grassmann even components of the traceless graded symmetric representation of $OSp(d|2)$ of dimension $\Delta$ and spin $\ell$, transforming under $SO(d)$.

The equivalence between the four-point superconformal block and the $(d-2)$ dimensional conformal block can be extended to higher-point blocks as well. 
Following the work of ref.~\cite{Kaviraj:2019tbg} for the four-point block, the proof by conformal Casimir equation of the equivalence of higher-point blocks should be for the most part automatic.
From the representation theory standpoint, one now needs to consider direct product representations over all exchanged supermultiplets in the intermediate channels. 
Viewing the direct product in components, instead of five terms as in~\eno{Rych4Ell}, one gets in total $5^{n-3}$ terms for an $n$-point block.
For instance, if we write~\eno{Rych4Ell} schematically as:
\eqn{}{
(d-2)_{(0,0)} \sim (d)_{(0,0)} + (d)_{(2,0)} + (d)_{(1,-1)} + (d)_{(0,-2)} + (d)_{(2,-2)}\,,
}
where the subscripts denote the shifts in the exchanged dimension and spin, then the dimensional reduction for the five-point block features 25 terms and takes the form
\eqn{}{
(d-2)_{(0,0),(0,0)} &\sim (d)_{(0,0),(0,0)} + (d)_{(0,0),(2,0)} + (d)_{(0,0),(1,-1)} + (d)_{(0,0),(0,-2)} + (d)_{(0,0),(2,-2)} \cr 
&+ (d)_{(2,0),(0,0)} + (d)_{(2,0),(2,0)} + (d)_{(2,0),(1,-1)} + (d)_{(2,0),(0,-2)} + (d)_{(2,0),(2,-2)} \cr 
&+ (d)_{(1,-1),(0,0)} + \cdots + (d)_{(1,-1),(2,-2)} \cr 
&+ (d)_{(0,-2),(0,0)} + \cdots + (d)_{(0,-2),(2,-2)} \cr 
&+ (d)_{(2,-2),(0,0)} + \cdots + (d)_{(2,-2),(2,-2)}\,,
}
where we now have two pairs of subscripts corresponding to the two intermediate exchange operators.
The precise coefficients, the analogs of $c_{i,j}$s in~\eno{Rych4Ell}, which make the  dimensional reduction for an $n$-point block of arbitrary topology possible remain non-trivial to determine.

In this paper we will be restricting focus to conformal blocks with scalar exchanges. 
For the four-point block it turns out when we set $\ell=0$ several coefficients in~\eno{Rych4Ell} vanish, such that the scalar exchange sector in fact {\it decouples} and yields a simpler dimensional reduction~\cite{Kaviraj:2019tbg}
\eqn{}{
(d-2)_{(0,-)} \sim (d)_{(0,-)} + (d)_{(2,-)} \,.
}
In the scalar exchange sector, as expected from direct product representations, the $n$-point block dimensional reduction will now involve $2^{n-3}$ terms.
For instance for the five-point block, we will have
\eqn{FiveSchem}{
(d-2)_{(0,-),(0,-)} \sim (d)_{(0,-),(0,-)} + (d)_{(0,-),(2,-)} 
+ (d)_{(2,-),(0,-)} + (d)_{(2,-),(2,-)} \,.
}
 An alternate way to motivate the form of higher-point dimensional reduction (in the scalar exchange sector) is via the dimensional reduction relations obeyed by four-point conformal partial waves~\cite{Zhou:2020ptb}. 
 Conformal partial waves admit a simple, iterative integral prescription for constructing higher-point partial waves from lower-point ones.
 Thus starting from the four-point dimensional reduction, one can iteratively construct the five-point and higher dimensional reductions. 
 The relations take the same form as in the case of conformal blocks but the precise coefficients appearing in the relations differ.
 The fact that the conformal partial waves obey dimensional reductions then suggests that conformal blocks might too.
 
Returning back to the dimensional reduction of an arbitrary $n$-point conformal block with scalar exchanges,
determining the coefficients in general which make the dimensional reductions such as~\eno{FiveSchem} possible remains a non-trivial challenge for two reasons. Firstly, these coefficients are topology dependent and the number of inequivalent conformal block topologies grows rapidly with $n$. Secondly, the number of terms ($2^{n-3}$) in each relation itself grows rapidly with $n$. 

  Notwithstanding, recent work on Feynman rules for higher-point conformal blocks~\cite{Hoback:2020pgj} will allow us in this paper to establish these relations and obtain the exact coefficients. 
  Notably, in this paper we will derive a remarkably simple set of Feynman-like rules for writing down the coefficients which appear in the dimensional reduction of any scalar conformal block of any topology  with scalar exchanges.
  The novelty of this result is the existence of a compact set of vertex and edge factors which can be used as building blocks to generate arbitrary coefficients for arbitrary conformal blocks. 
    Interestingly, we will be able to determine the full set of rules (i.e.\ vertex and edge factors) just by looking at three specific conformal blocks: the four-point block, the five-point block and the six-point block in the OPE/snowflake channel.
  These rules are suggestive of the existence of, and pave the way for determining similar rules for the more general higher-point dimensional reduction relations which admit spinning exchanges.

\vspace{0.75em}
The outline for the rest of the paper is as follows:
In section~\ref{DIMREDFOUR} we begin by reviewing the four-point dimensional reduction, and in section~\ref{DIMREDHIGHER} we introduce a convenient graphical notation for writing down precise dimensional reduction relations for any scalar block. 
In particular we propose a set of Feynman-like rules for working out every coefficient in the dimensional reduction of an arbitrary $n$-point block, which we proceed to prove in section~\ref{FEYNMAN}.
We provide various consistency checks of our results, including consistency with the OPE limit and the unit operator limit in section~\ref{OPEUNIT}.
We conclude with final remarks and future directions in section~\ref{DISCUSS}.
In appendix~\ref{PARTIALWAVES} we expand on our comments above on the dimensional reduction of higher-point conformal partial waves.
In appendix~\ref{HOGER} we revisit earlier work on dimensional reduction of four-point conformal blocks~\cite{Hogervorst:2016hal} in light of the (finite term) dimensional reduction relations of ref.~\cite{Kaviraj:2019tbg}, and in appendix~\ref{APP:EXAMPLES} we present some additional illustrative examples of higher-point dimensional reduction relations and list various topologies on which we tested our results as non-trivial consistency checks.

\section{Dimensional reduction of higher-point blocks}
\label{DIMRED}

In this section we begin by reviewing the relevant dimensional reduction results for four-point conformal blocks. 
Subsequently, we will present the generalization to all higher-point conformal blocks.
Finally, we will prove this novel proposal with help of the recently proposed Feynman rules for constructing higher-point scalar conformal blocks in arbitrary topologies, and end with various consistency checks.

\subsection{Review of four-point dimensional reduction}
\label{DIMREDFOUR}

Using conformal invariance, one can write the four-point function of scalar primary operators as
\eqn{4ptfn}{
\langle {\cal O}_1(x_1) {\cal O}_2(x_2) {\cal O}_3(x_3) {\cal O}_4(x_4) \rangle = W_{0}(x_i)\:  f(u,v)
}
where the kinematic {\it leg-factor} $W_{0}(x_i)$ accounts for the correct behaviour of the four-point function under conformal transformations and is given by 
\eqn{W0Def}{
W_{0}(x_i) 
 &=  {\left({x_{24}^2 \over x_{14}^2 }\right)^{\Delta_1-\Delta_2\over 2}  \left({x_{14}^2 \over x_{13}^2 }\right)^{\Delta_3-\Delta_4\over 2}  \over (x_{12}^2)^{\Delta_1+\Delta_2 \over 2} (x_{34}^2)^{\Delta_3 + \Delta_4 \over 2} }\,,
}
whereas $f(u,v)$ is an undetermined function of conformal cross-ratios
\eqn{uvDef}{
u = {x_{12}^2 x_{34}^2 \over x_{13}^2 x_{24}^2} \qquad \qquad v = {x_{14}^2 x_{23}^2 \over x_{13}^2 x_{24}^2}  \,,
}
which captures the dynamics of the theory. It takes the general form,
\eqn{fDef}{
f(u,v) = \sum_{\Delta, \ell} c_{12{\cal O}} c_{34 {\cal O}} \: g_{\Delta, \ell}(u,v)\,,
}
where the sum is over primaries ${\cal O}$ with dimension $\Delta$ and spin $\ell$, which appear in the spectrum of the CFT, and the $c_{ijk}$s are the associated OPE coefficients. 
The expansion above is organized such that the function $g_{\Delta,\ell}$ packages the contribution to the four-point function originating from the exchange of the highest-weight representation labeled by $(\Delta, \ell)$ and its full conformal family in the $(12)(34)$ exchange channel.

We refer to $g_{\Delta,\ell}$ as the {\it bare conformal block}, which is proportional to the {\it conformal block} $W_{\Delta,\ell}(x_i) = W_0(x_i)\: g_{\Delta,\ell}(u,v)$ via the leg factor.\footnote{Sometimes it is also useful to think of the four-point conformal block as an eigenfunction of the quadratic conformal Casimir operator with appropriate eigenvalues and boundary conditions~\cite{Dolan:2000ut,Dolan:2003hv}.}
Diagrammatically, we represent the block as shown in figure~\ref{fig:FourCB}.
\begin{figure}[htb]
    \centering
    \[ W_{\Delta,\ell} = W_0 \: g_{\Delta,\ell} = \musepic{\figFourCB} \]
    \caption{Scalar four-point conformal block. }
    \label{fig:FourCB}
\end{figure}
For the remainder of this paper we will include a superscript on the (bare) conformal block to denote the spacetime dimension of the CFT, that is from here on we will write $W^{(d)}_{\Delta,\ell} = W_0 g_{\Delta,\ell}^{(d)}$, where for brevity we have suppressed the coordinate/cross-ratio dependence.

As discussed in the introduction, the four-point block satisfies a beautiful dimensional reduction relation with a finite number of terms~\cite{Kaviraj:2019tbg}.
When the bare blocks are normalized using the normalization conventions of refs.~\cite{Dolan:2000ut,Dolan:2003hv} this relation takes the form~\eno{Rych4Ell}, repeated below for convenience
\eqn{4RychEll}{
g_{\Delta,\ell}^{(d-2)} = g_{\Delta,\ell}^{(d)} + c_{2,0}\: g_{\Delta + 2,\ell}^{(d)} + c_{1,-1}\: g_{\Delta+1,\ell-1}^{(d)} + c_{0,-2}\: g_{\Delta,\ell-2}^{(d)} + c_{2,-2}\: g_{\Delta+2,\ell-2}^{(d)} \,,
}
where $c_{i,j}$s are known constants that the depend on quantities such as $\Delta_1-\Delta_2, \Delta_3-\Delta_4, \Delta, \ell$ and spacetime dimension $d$.
For example,
\eqn{c20EllDef}{
c_{2,0} =   { -(\Delta-1)\Delta \left({\Delta+\Delta_1-\Delta_2 +\ell \over 2} \: {\Delta+\Delta_2-\Delta_1 +\ell \over 2} \: {\Delta+\Delta_3-\Delta_4 +\ell \over 2} \: {\Delta+\Delta_4-\Delta_3 + \ell \over 2} \right) \over (\Delta+\ell-1) (\Delta+\ell)^2 (\Delta_{\delta_1}+ \ell +1) (\Delta_{\delta_1}-{d\over 2}+1)  (\Delta_{\delta_1}-{d\over 2}+2)} \,.
}
Since the conformal block is proportional to the bare block via the kinematic factor in~\eno{W0Def}, the blocks themselves satisfy the relation~\eno{4RychEll}. 

The relation~\eno{4RychEll} is surprising from two points of view. 
When dimensionally reducing a $d$-dimensional CFT (say, to a $d^\prime < d$ dimensional theory), each multiplet of $SO(d+1,1)$ rearranges into an infinite family of multiplets of $SO(d^\prime+1,1)$~\cite{Hogervorst:2016hal}.\footnote{Generically, the CFT in $d^\prime$ dimensions is non-local with no conserved stress-tensor.}
Thus the four-point block in $d$ dimensions splits into an infinite series of $d^\prime$-dimensional blocks.
Generically, one would expect that inverting this relation to express $d^\prime$-dimensional blocks in terms of $d$-dimensional ones would yield an infinite series as well. 
Surprisingly when $d-d^\prime$ is even, this is not the case. 
This observation came as a result of a precise relation between Parisi-Sourlas SCFTs in $d$ dimensions and a local non-supersymmetric CFT in $d-2$ dimensions~\cite{Parisi:1979ka,Kaviraj:2019tbg}, which implies that there exists a one-to-one correspondence between certain multiplets of the SCFT and the dimensionally reduced CFT~\cite{Kaviraj:2019tbg}.
In hindsight, from the point of view of the infinite series expansion of ref.~\cite{Hogervorst:2016hal}, one can operationally attribute the finite term inverse relation to the fact that the series coefficients satisfy remarkable recursion relations. 
These recursion relations are responsible for miraculous cancellations that allow us to invert the infinite series expansion to obtain a finite series. 
In appendix ~\ref{HOGER}, we present these recursion relations and show that using these relations one obtains a finite series upon inverting the infinite series expansion.

Furthermore, Dolan and Osborn found an interesting relation between $d$-dimensional and $(d-2)$-dimensional four-point blocks~\cite[eq.\ (5.4)]{Dolan:2003hv}, which looks similar to~\eno{4RychEll} but is distinct in important ways. 
For one, the Dolan-Osborn relation provides a dimensional {\it uplift}, which produces a representation of the $d$-dimensional block in terms of a linear combination of five $(d-2)$-dimensional ones, with the dimension and spin of the intermediate operator shifted in the opposite direction.
The opposite sign shifts are expected for dimensional uplifts.
However, in contrast to the dimensional reduction, the linear combination of $(d-2)$-dimensional blocks gives the $d$-dimensional block up to an overall factor of cross-ratios.
Moreover, in the dimensional reduction relation the scalar exchange sector \textit{decouples} as noted in section~\ref{INTRO}.
However, the scalar exchange sector does not decouple in the dimensional uplift relation.\footnote{Notwithstanding, it turns out it is still possible to obtain~\eno{4RychEll} starting from various identities and relations of ref.~\cite{Dolan:2003hv}.}

In this paper we will focus on the scalar exchange sector mentioned above. When $\ell=0$, the  coefficients in the final three terms of~\eno{4RychEll} vanish to give the simplified relation
\eqn{Rych4scalar}{
g^{(d-2)}_{\Delta_{\delta_1}} =  g^{(d)}_{\Delta_{\delta_1}} + 
{ -\Delta_{\delta_11,2} \: \Delta_{\delta_12,1} \: \Delta_{\delta_13,4} \: \Delta_{\delta_14,3} \over \Delta_{\delta_1} (\Delta_{\delta_1}+1) (\Delta_{\delta_1}-{d\over 2}+1)  (\Delta_{\delta_1}-{d\over 2}+2)}\: 
g^{(d)}_{\Delta_{\delta_1}+2}\,,
}
where we have suppressed the $\ell=0$ subscript in the bare blocks and  have relabeled the exchanged conformal dimension $\Delta_{\delta_1}$.
Note that spinning exchanges have completely dropped out from the dimensional reduction.\footnote{This is also expected from the Parisi-Sourlas SCFT representation theory point of view~\cite{Kaviraj:2019tbg}.}
Here and below we will be using the following shorthand for conformal dimensions:
\eqn{Deltaijk}{
\Delta_{i_1\ldots i_{j},i_{j+1} \ldots i_k} := {\Delta_{i_1} + \cdots + \Delta_{i_j} - \Delta_{i_{j+1}} - \cdots -\Delta_{i_k} \over 2}\,.
}

\subsection{Higher-point dimensional reduction}
\label{DIMREDHIGHER}

For generalization of~\eno{Rych4scalar} to higher-point blocks of arbitrary topology, it is convenient to introduce a graphical notation for the coefficients. For example, we rewrite~\eno{Rych4scalar} as 
\eqn{Rych4fig}{
g^{(d-2)}_{\Delta_{\delta_1}} = \figCFour{\Delta_1}{\Delta_2}{\Delta_3}{\Delta_4}{\Delta_{\delta_1}}{}  g^{(d)}_{\Delta_{\delta_1}} + \figCFour{\Delta_1}{\Delta_2}{\Delta_3}{\Delta_4}{\Delta_{\delta_1}}{1}  g^{(d)}_{\Delta_{\delta_1}+2}\,.
}
The coefficients above are represented by shaded rectangular boxes depicting the topology of the conformal block. All internal vertices are shown as blue dots, and for terms where the internal dimension of the block is shifted up by 2 (e.g.\ the second term of~\eno{Rych4fig}), the corresponding internal edge in the graphical representation of the coefficient is distinguished with a dashed-red line.
Of course, comparing~\eno{Rych4fig} with~\eno{Rych4scalar} we have
\eqn{FourCoeffs}{
\figCFour{\Delta_1}{\Delta_2}{\Delta_3}{\Delta_4}{\Delta_{\delta_1}}{} = 1\,, \quad 
\figCFour{\Delta_1}{\Delta_2}{\Delta_3}{\Delta_4}{\Delta_{\delta_1}}{1} = { -\Delta_{\delta_11,2} \: \Delta_{\delta_12,1} \: \Delta_{\delta_13,4} \: \Delta_{\delta_14,3} \over \Delta_{\delta_1} (\Delta_{\delta_1}+1) (\Delta_{\delta_1}-{d\over 2}+1)  (\Delta_{\delta_1}-{d\over 2}+2)} \,.
}

It turns out the graphical notation admits a natural generalization to all higher-point blocks in arbitrary topologies.
One can write down a set of graphical rules for reconstructing the mathematical expression associated with each graphical coefficient appearing in dimensional reduction relations.
These rules associate a simple multiplicative factor to each internal vertex and edge in the graphical coefficient. 
The full coefficient is obtained by taking a product over all associated vertex and edge factors. 

The precise factors take the following form:
To each internal vertex, depending on how many incident edges are dashed-red, assign the corresponding factor of: 
\eqn{VertexRules}{
\begin{gathered}
\DrawVertex{\Delta_1}{}{\Delta_2}{}{\Delta_3}{} = 1\,,\qquad \qquad
\DrawVertex{\Delta_1}{}{\Delta_2}{}{\Delta_3}{1} = -\Delta_{31,2} \: \Delta_{32,1} \cr  
\DrawVertex{\Delta_1}{1}{\Delta_2}{1}{\Delta_3}{} =  -\Delta_{12,3}\: (\Delta_{12,3}+1) \left(\Delta_{123,}-{d\over 2}+1\right)  \left(\Delta_{12,3}-{d\over 2}+2\right) \cr  
\DrawVertex{\Delta_1}{1}{\Delta_2}{1}{\Delta_3}{1} = \Delta_{12,3}\: \Delta_{23,1} \: \Delta_{31,2} \left(\Delta_{123,} - {d\over 2} +1\right) \left(\Delta_{123,} -{d\over 2} +2\right) \left(\Delta_{123,} - d +3\right) .
\end{gathered}
}
To each internal edge assign a factor of
\eqn{EdgeRules}{
\DrawEdge{\Delta_{\delta_1}}{} = 1 \qquad \qquad \DrawEdge{\Delta_{\delta_1}}{1} = {-1 \over  \Delta_{\delta_1} (\Delta_{\delta_1}+1) (\Delta_{\delta_1}-{d\over 2}+1)  (\Delta_{\delta_1}-{d\over 2}+2) } \,,
}
depending on whether the edge is dashed-red or not.
The origin of these rules will be discussed in the next subsection. 

Applying these rules to the four-point coefficients in~\eno{Rych4fig}, we get
\eqn{}{
\figCFour{\Delta_1}{\Delta_2}{\Delta_3}{\Delta_4}{\Delta_{\delta_1}}{} &= \DrawVertex{\Delta_1}{}{\Delta_2}{}{\Delta_{\delta_1}}{} \DrawVertex{\Delta_3}{}{\Delta_4}{}{\Delta_{\delta_1}}{}
\DrawEdge{\Delta_{\delta_1}}{} =1\cr 
\figCFour{\Delta_1}{\Delta_2}{\Delta_3}{\Delta_4}{\Delta_{\delta_1}}{1} &= \DrawVertex{\Delta_1}{}{\Delta_2}{}{\Delta_{\delta_1}}{1} \DrawVertex{\Delta_3}{}{\Delta_4}{}{\Delta_{\delta_1}}{1}
\DrawEdge{\Delta_{\delta_1}}{1} = { -\Delta_{\delta_11,2} \: \Delta_{\delta_12,1} \: \Delta_{\delta_13,4} \: \Delta_{\delta_14,3} \over \Delta_{\delta_1} (\Delta_{\delta_1}+1) (\Delta_{\delta_1}-{d\over 2}+1)  (\Delta_{\delta_1}-{d\over 2}+2)}\,,
}
which reproduces~\eno{FourCoeffs} as desired.
For illustrative purposes, we also present a higher-point example, which will later appear as a coefficient in the dimensional reduction of the six-point block in the OPE/snowflake channel:
\eqn{6OPEcoeff}{
\figCSixOPE{\Delta_1}{\Delta_2}{\Delta_3}{\Delta_4}{\Delta_5}{\Delta_6}{\Delta_{\delta_1}}{}{\Delta_{\delta_2}}{1}{\Delta_{\delta_3}}{1} 
&= \DrawVertex{\Delta_1}{}{\Delta_2}{}{\Delta_{\delta_1}}{} 
\DrawVertex{\Delta_3}{}{\Delta_4}{}{\Delta_{\delta_2}}{1}
\DrawVertex{\Delta_5}{}{\Delta_6}{}{\Delta_{\delta_3}}{1}
\DrawVertex{\Delta_{\delta_2}}{1}{\Delta_{\delta_3}}{1}{\Delta_{\delta_1}}{}
\DrawEdge{\Delta_{\delta_1}}{}
\DrawEdge{\Delta_{\delta_2}}{1}
\DrawEdge{\Delta_{\delta_3}}{1} \cr 
&= { -\Delta_{\delta_23,4} \: \Delta_{\delta_24,3} \: \Delta_{\delta_35,6} \: \Delta_{\delta_36,5} 
\Delta_{\delta_2\delta_3,\delta_1}
\over \prod_{i=2}^3  \Delta_{\delta_i} (\Delta_{\delta_i}+1) (\Delta_{\delta_i}-{d\over 2}+1)  (\Delta_{\delta_i}-{d\over 2}+2) } \cr 
&\quad \times (\Delta_{\delta_2\delta_3,\delta_1}+1) \left(\Delta_{\delta_2\delta_3\delta_1,}-{d\over 2}+1\right)  \left(\Delta_{\delta_2\delta_3,\delta_1}-{d\over 2}+2\right).
}

\begin{figure}[!tb]
    \centering
    \[ W_{\Delta_{\delta_1},\Delta_{\delta_2}}  = \musepic{\figFiveCB} \]
    \caption{Five-point conformal block with scalar exchanges. }
    \label{fig:FiveCB}
\end{figure}

Setting up the graphical notation merely for the four-point example is admittedly overkill, but
as mentioned previously it affords significant payoff at higher points. 
For example, the scalar five-point block with scalar exchanges shown in figure~\ref{fig:FiveCB}  satisfies an elegant dimensional reduction, whose schematic form was presented in~\eno{FiveSchem},
\eqn{Rych5}{
g^{(d-2)}_{\Delta_{\delta_1},\Delta_{\delta_2}} &= \figCFive{\Delta_1}{\Delta_2}{\Delta_3}{\Delta_4}{\Delta_5}{\Delta_{\delta_1}}{}{\Delta_{\delta_2}}{} g^{(d)}_{\Delta_{\delta_1},\Delta_{\delta_2}} 
+ \figCFive{\Delta_1}{\Delta_2}{\Delta_3}{\Delta_4}{\Delta_5}{\Delta_{\delta_1}}{1}{\Delta_{\delta_2}}{} g^{(d)}_{\Delta_{\delta_1}+2,\Delta_{\delta_2}} \cr 
&+ \figCFive{\Delta_1}{\Delta_2}{\Delta_3}{\Delta_4}{\Delta_5}{\Delta_{\delta_1}}{}{\Delta_{\delta_2}}{1} g^{(d)}_{\Delta_{\delta_1},\Delta_{\delta_2}+2} 
+ \figCFive{\Delta_1}{\Delta_2}{\Delta_3}{\Delta_4}{\Delta_5}{\Delta_{\delta_1}}{1}{\Delta_{\delta_2}}{1} g^{(d)}_{\Delta_{\delta_1}+2,\Delta_{\delta_2}+2} \,,
}
where the coefficients are determined using the simple rules stated above.\footnote{Just like in the four-point case, the full five-point conformal block also satisfies the same relation since it is proportional to the bare block up to known kinematic leg-factors.}

The general pattern should hopefully be evident from the four- and five-point examples~\eno{Rych4fig} and~\eno{Rych5}: 
\begin{quote}  
\fbox{\parbox{0.9\textwidth}{For an $n$-point block of a given topology, the $(d-2)$-dimensional block will be given by a linear combination of $d$-dimensional blocks of the same topology and same external dimensions, where each of the $(n-3)$ internal dimensions is either shifted up by two or remains unchanged. 
This gives rise to $2^{n-3}$ possible combinations, and each of these appear in the linear combination weighted by a constant coefficient. This coefficient is given graphically by an unrooted binary tree representation of the conformal block with the edges corresponding to all shifted internal dimensions shaded dashed-red. 
The coefficient factorizes into a product over its constituent internal vertex and edge factors which are given by~\eno{VertexRules}-\eno{EdgeRules}.}}
\end{quote}
These dimensional relations hold  both  for the bare blocks as well as for the full conformal blocks.

\begin{figure}[!htb]
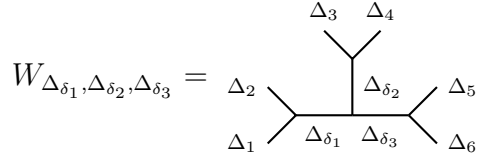

    \centering
    \[ W_{\Delta_{\delta_1},\Delta_{\delta_2},\Delta_{\delta_3}}  = \musepic{\figSixOPECB} \]
    \caption{Six-point conformal block in OPE/snowflake channel. }
    \label{fig:SixOPECB}
\end{figure}

For example, the six-point block in the  OPE/snowflake channel shown in figure~\ref{fig:SixOPECB}, for which an explicit power series representation in $d$ dimensions was first obtained in ref.~\cite{Fortin:2020yjz},  satisfies:
\begingroup
\allowdisplaybreaks
\begin{align*}
& g^{(d-2)}_{\Delta_{\delta_1},\Delta_{\delta_2},\Delta_{\delta_3}} \\
&= \figCSixOPE{\Delta_1}{\Delta_2}{\Delta_3}{\Delta_4}{\Delta_5}{\Delta_6}{\Delta_{\delta_1}}{}{\Delta_{\delta_2}}{}{\Delta_{\delta_3}}{} 
g^{(d)}_{\Delta_{\delta_1},\Delta_{\delta_2},\Delta_{\delta_3}} 
+ \figCSixOPE{\Delta_1}{\Delta_2}{\Delta_3}{\Delta_4}{\Delta_5}{\Delta_6}{\Delta_{\delta_1}}{1}{\Delta_{\delta_2}}{}{\Delta_{\delta_3}}{} 
g^{(d)}_{\Delta_{\delta_1}+2,\Delta_{\delta_2},\Delta_{\delta_3}} \\ 
&+ \figCSixOPE{\Delta_1}{\Delta_2}{\Delta_3}{\Delta_4}{\Delta_5}{\Delta_6}{\Delta_{\delta_1}}{}{\Delta_{\delta_2}}{1}{\Delta_{\delta_3}}{} 
g^{(d)}_{\Delta_{\delta_1},\Delta_{\delta_2}+2,\Delta_{\delta_3}}  
 + \figCSixOPE{\Delta_1}{\Delta_2}{\Delta_3}{\Delta_4}{\Delta_5}{\Delta_6}{\Delta_{\delta_1}}{}{\Delta_{\delta_2}}{}{\Delta_{\delta_3}}{1} 
 g^{(d)}_{\Delta_{\delta_1},\Delta_{\delta_2},\Delta_{\delta_3}+2}  \\ 
& +\figCSixOPE{\Delta_1}{\Delta_2}{\Delta_3}{\Delta_4}{\Delta_5}{\Delta_6}{\Delta_{\delta_1}}{1}{\Delta_{\delta_2}}{1}{\Delta_{\delta_3}}{}
g^{(d)}_{\Delta_{\delta_1}+2,\Delta_{\delta_2}+2,\Delta_{\delta_3}} 
+ \figCSixOPE{\Delta_1}{\Delta_2}{\Delta_3}{\Delta_4}{\Delta_5}{\Delta_6}{\Delta_{\delta_1}}{}{\Delta_{\delta_2}}{1}{\Delta_{\delta_3}}{1}
g^{(d)}_{\Delta_{\delta_1},\Delta_{\delta_2}+2,\Delta_{\delta_3}+2} \\ 
& + \figCSixOPE{\Delta_1}{\Delta_2}{\Delta_3}{\Delta_4}{\Delta_5}{\Delta_6}{\Delta_{\delta_1}}{1}{\Delta_{\delta_2}}{}{\Delta_{\delta_3}}{1} 
g^{(d)}_{\Delta_{\delta_1}+2,\Delta_{\delta_2},\Delta_{\delta_3}+2} 
+ \figCSixOPE{\Delta_1}{\Delta_2}{\Delta_3}{\Delta_4}{\Delta_5}{\Delta_6}{\Delta_{\delta_1}}{1}{\Delta_{\delta_2}}{1}{\Delta_{\delta_3}}{1}
g^{(d)}_{\Delta_{\delta_1}+2,\Delta_{\delta_2}+2,\Delta_{\delta_3}+2}\,.
\stepcounter{equation}\tag{\theequation}\label{Rych6OPE}
\end{align*}
\endgroup
The constant coefficients are determined using the rules above; one of them was worked out in~\eno{6OPEcoeff}.

For the reader's reference, a few more explicit higher-point examples are presented in appendix~\ref{APP:EXAMPLES}.
We will now prove the rules for obtaining the coefficients of the dimensional reduction.

\subsection{Determining the higher-point coefficients}
\label{FEYNMAN}

 As noted previously, Parisi-Sourlas theory suggests the following general form for the dimensional reduction of an $n$-point conformal block of a fixed topology (restricted to scalar exchanges),
\eqn{ToFind}{
g^{(d-2)}_{\Delta_{\delta_1},\ldots,\Delta_{\delta_i}, \ldots, \Delta_{\delta_{n-3}}} = \left(\prod_{i=1}^{n-3}\sum_{m_i \in \{0,1\}}\right) \lambda_{m_1 m_2 \ldots m_{n-3}}^{(t)} \: g^{(d)}_{\Delta_{\delta_1} + 2m_1,\ldots,\Delta_{\delta_i}+2m_i, \ldots, \Delta_{\delta_{n-3}}+2m_{n-3}} \,,
}
where the $2^{n-3}$ constants $\lambda_{m_1 \ldots m_{n-3}}^{(t)}$ are to be determined. 
The superscript $(t)$ on the constants emphasizes the dependence on the topology of the block.
Naively, determining these constants for {\it every} $n$-point topology appears to be a hopeless task, even if the conformal blocks entering into~\eno{ToFind} are known, since the number of inequivalent topologies grows rapidly with $n$.\footnote{There are no known closed-form expressions for the number of inequivalent topologies; see the integer sequence~\cite{oeis} for the first few terms in the sequence as well as an iterative prescription.}
 
A similar obstacle arose in a related context~\cite{Fortin:2020bfq,Hoback:2020pgj}. 
Until recently, only a few isolated examples of  $d$-dimensional higher-point conformal blocks were known, in terms of power series representations.
These representations were worked out one block or topology at a time.
However for the eventual goal of finding all blocks, working them out one topology at a time is both inefficient and impractical.
To get around this, a more attractive approach was conjectured in the form of a set of constructive ``Feynman rules''~\cite{Hoback:2020pgj}, similar to Feynman rules for Mellin amplitudes~\cite{Fitzpatrick:2011ia,Paulos:2011ie}.\footnote{See also ref.~\cite{Fortin:2020bfq}. The  conjectured rules were later proven rigorously in ref.~\cite{InPrep}.}
These rules furnish a power series representation of any conformal block in terms of {\it local} graph theoretic data (such as vertex and edge factors), to be reviewed below.
This approach provides a simple and straightforward way to write down an explict series representation of any conformal block in any topology  involving scalar exchanges directly from its binary tree representation.

The previous discussion suggests a method for determining the constants in~\eno{ToFind}.
One can substitute the power series expansions of the blocks prescribed by the Feynman rules on both sides of~\eno{ToFind} and hope to solve for the undetermined coefficients.
Notably, due to the fact that the conformal blocks in~\eno{ToFind} are specified by local graph theoretic Feynman rules, the coefficients which solve~\eno{ToFind} themselves turn out to obey local Feynman-like rules with their own (finite) set of yet-to-be-determined vertex and edge factors.
It turns out it is sufficient to explicitly work out a handful of cases to completely fix all vertex and edge factors once and for all. Having fixed the factors these rules yield the dimensional reduction of {\it all} higher-point blocks.
These are precisely the rules proposed in~\eno{VertexRules}-\eno{EdgeRules} in the form of a compact set of vertex and edge factors.
We will now show how to obtain these factors in the remainder of this section. 
Before proceeding we will briefly review the most pertinent aspects of the aforementioned Feynman rules for conformal blocks~\cite{Hoback:2020pgj}.
Subsequently, we will use these rules to obtain the vertex and edge factors.

\vspace{0.75em}

Any $n$-point scalar conformal block with scalar exchanges labeled by dimensions $\Delta_{\delta_i}$ in any given topology takes the following form,
\eqn{}{
 W(x_i) = W_0(x_i) \: g^{(d)}(u,1-v)\,,
 }
 where $W_0$ is the $n$-point kinematic leg-factor which accounts for the appropriate scaling and behavior under conformal transformations, and the bare block takes the form of a power series in powers of cross-ratios:\footnote{A notational remark: In ref.~\cite{Hoback:2020pgj}, the function $g(u,1-v)$ refers to~\eno{bareblockDef} without the overall factor of $ \left(\prod_{i=1}^{n-3} u_i^{\Delta_{\delta_i}/2} \right)$.}
\eqn{bareblockDef}{
  g^{(d)}(u,1-v) =  \left(\prod_{i=1}^{n-3} u_i^{\Delta_{\delta_i}/2} \right)\sum_{k_i, j_{rs}=0}^\infty \left[\left( \prod_{i=1}^{n-3} {u_i^{k_i} \over k_i!} \right) \left(\prod_{(rs)}^{\binom{n-2}{2}} {\left(1-v_{rs}\right)^{j_{rs}} \over j_{rs}!} \right) \left( \prod_{i=1}^{n-3} E_i \right) \left(\prod_{i=1}^{n-2} V_i \right) \right]
 }
where $u_i, v_{rs}$ form a set of appropriate $n(n-3)/2$ independent cross-ratios. The precise prescription for the cross-ratios is not important here -- it is sufficient to note that there exists one which admits an expansion of the form~\eno{bareblockDef}~\cite{InPrep}.

To understand the main ingredients which go into~\eno{bareblockDef}, first note that each $n$-point conformal block has a unique unrooted binary tree representation with $n$ leaves which comes with $(n-3)$ internal edges and $(n-2)$ internal vertices.
The internal edges and internal vertices of the binary tree representation correspond one-to-one with $(n-3)$ edge-factors $E_i$ and $(n-2)$ vertex factors $V_i$ in~\eno{bareblockDef}, respectively.
Each internal line carrying the dimension $\Delta_{\delta_i}$ is associated with a unique integral parameter $k_i$, while $k_i = 0$ for all external legs. 
The edge factor associated to each internal line is then given by
\eqn{EdgeDef}{
     E_i := { (\Delta_{\delta_i} - h +1)_{k_i} \over (\Delta_{\delta_i})_{2k_i + \ell_{\delta_i}} }\,,
}
where $h := d/2$ and $(a)_b := \Gamma(a+b)/\Gamma(a)$.
For a given vertex, the vertex factor $V_i$ is determined by looking at the binary tree representation and determining the dimension labels on the three edges incident on the vertex.
Let these dimensions be $\Delta_a, \Delta_b$, and $\Delta_c$ and let the associated parameters for the edges be $k_a, k_b$, and $k_c$. Then the corresponding vertex factor is given by
\eqn{VertexDef}{
V_i &:= (\Delta_{ab,c})_{k_{ab,c}+{1\over 2}\ell_{ab,c}} (\Delta_{ac,b})_{k_{ac,b}+{1\over 2}\ell_{ac,b}} (\Delta_{bc,a})_{k_{bc,a}+{1\over 2}\ell_{bc,a}} \cr 
     &  \times F_A^{(3)}\!\left[\Delta_{abc,}- h; \{-k_a, -k_b, -k_c \}; \left\{\Delta_a -h+1, \Delta_b -h+1,\Delta_c -h+1 \right\}; 1,1,1 \right] ,
}
where $F_A^{(3)}$ is the Lauricella function of three variables.\footnote{The Lauricella function is a multi-variable generalized hypergeometric function  defined via
    \eqn{LauricellaDef}{
    F_A^{(\ell)}\Big[g;\{a_1,\ldots,a_\ell\};\{b_1,\ldots,b_\ell\};x_1,\ldots,x_\ell\Big]
    := \left[\prod_{i=1}^\ell\sum_{n_i=0}^\infty \right](g)_{\sum_{i=1}^\ell n_i}\prod_{i=1}^\ell \frac{(a_i)_{n_i}}{(b_i)_{n_i}}\frac{x_i^{n_i}}{n_i!}\,.
    }
See ref.~\cite[appendix A]{Hoback:2020pgj} for its relation to other familiar functions when all its arguments $x_i=1$.}
Here we are using the convention,
\eqn{kellijk}{
    k_{i_1 \ldots i_m, i_{m+1} \ldots i_{n}} &:= k_{i_1} + \cdots + k_{i_m} - k_{i_{m+1}} - \cdots - k_{i_n}  \cr 
    \ell_{i_1 \ldots i_m, i_{m+1} \ldots i_{n}} &:= \ell_{i_1} + \cdots + \ell_{i_m} - \ell_{i_{m+1}} - \cdots - \ell_{i_n} \,,
}
which is notably missing an overall factor of half compared to the convention for conformal dimensions~\eno{Deltaijk}.
The precise knowledge of the $\ell_{a}$ symbols will be unnecessary for our purposes and it suffices to say they are given by positive linear combinations of the dummy parameters $j_{rs}$ and can be determined iteratively from the local graph structure of the binary tree representation.  
Note that in total three types of (cubic) vertices are possible: one with all incident edges internal, one with exactly two incident edges internal (and one external), or one with exactly one incident edge internal (and the remaining two external).
In the latter two cases the $k_a$ parameters associated to external legs are set to zero and the Lauricella function of three variables in~\eno{VertexDef} reduces to the Lauricella function of two or one variable(s), respectively.

\vspace{.75em}

Having reviewed the Feynman Rules for conformal blocks, we return back to the calculation of the coefficients in~\eno{ToFind} for an $n$-point conformal block of any topology. 
We start by substituting the expression for the blocks on the left and right hand sides of~\eno{ToFind} using the Feynman rules described above.
Note that the dependence on coordinates in~\eno{ToFind} is via the exponents of the cross-ratios in~\eno{bareblockDef}, which depend solely on the exchanged dimensions $\Delta_{\delta_i}$.  
The dependence of the blocks on the spacetime dimension $d=2h$ is only via the position-independent edge and vertex factors~\eno{EdgeDef}-\eno{VertexDef}.
Thus the shift $d \to d-2$ on the left hand side of~\eno{ToFind} does not formally affect the coordinate dependence.
On the right hand side since internal dimensions get shifted up by two in all possible combinations, one may naively expect the coordinate dependence in each term is altered.
However, any such integral shift in the internal dimensions $\Delta_{\delta_i}$  can be compensated for by shifting the associated dummy variables $k_i$ in~\eno{bareblockDef} down by one. 
This shift does not affect the limits of the $k_i$ sums in~\eno{bareblockDef} due to the factor of $k_i!$ in the denominator. 
Conveniently, it restores the overall scaling of the cross-ratios $u_i$ on the right hand side to make it manifestly match the scaling on the left hand side.
Then, bringing everything in~\eno{ToFind} to one side we set the coefficient at each order in powers of $\prod_i u_i^{k_i}$ to zero.
Doing this we obtain an infinite set of linear equations with {\it constant} coefficients, one for each combination of non-negative integral values for the set $\{k_1, k_2, k_3, \ldots, k_{n-3} \}$, which we can solve to obtain the  position-independent coefficients $\lambda^{(t)}_{\ldots}$.\footnote{In the calculation explained here, we need only focus on the $(n-3)$-dimensional sum over the $k_i$s in~\eno{bareblockDef};  the $\binom{n-2}{2}$-dimensional sum over the $j_{rs}$ parameters remains unaffected through the entire calculation and does not play any role.}

Since the conformal blocks are given by Feynman rules, each term in~\eno{ToFind} is given by a product of local vertex and edge factors depending on the topology and labeling of the binary graph representation of the block. 
Shifting the internal exchanged dimensions or the spacetime dimension rescales each of these factors appropriately. 
The  product over the rescaled factors resulting from these shifts maintains the local product structure of the Feynman rules. 
Consequently, the linear equations obtained above inherit the Feynman rules-like structure.
Thus their solution is given in terms of $\lambda_{\ldots}^{(t)}$ coefficients which themselves obey Feynman-like rules that are closely tied to the binary graph representations of the individual blocks on the right hand side of~\eno{ToFind}.

 The discussion above leads naturally to the following ansatz for $\lambda_{\ldots}^{(t)}$:
\begin{itemize}
    \item The coefficients are fully determined in terms of the associated (labeled) binary graph representation of the conformal blocks in~\eno{ToFind}. 
    
    Note that the topology of each conformal block appearing in~\eno{ToFind} is the same; the only feature distinguishing one block from another is the precise set of exchanged dimensions in the intermediate channels. 
    Thus the coefficient $\lambda_{\ldots}^{(t)}$ will be fully specified by its labeled binary graph representation, and for brevity, instead of labeling a {\it shifted} internal line with a dimension $\Delta_{\delta} +2$, we will draw all shifted intermediate legs as dashed-red lines and simply label them with the original unshifted dimension $\Delta_{\delta}$ in red. See e.g.\ \eno{Rych6OPE}.
    
    \item These coefficients satisfy a factorization property: They can be written as a product of local terms, one for each internal vertex and internal edge in the binary graph. See e.g.\ the first equality in~\eno{6OPEcoeff}.

\end{itemize}
Our task now is to determine these associated vertex and edge factors. Note that there are only four possible types of vertex factors and two types of edge factors which can appear in the binary graph representation of {\it any} higher-point conformal block with scalar exchanges.
In the graphical convention described above they are given by~\eno{VertexRules}-\eno{EdgeRules}.
Thus we will only need to consider a finite number of examples to completely fix them.

We begin with the relation~\eno{Rych4fig}, where the coefficients are already known to be given by~\eno{FourCoeffs}~\cite{Kaviraj:2019tbg}.
Imposing the ansatz above yields the following constraints: 
\eqn{FourRqr}{
\DrawVertex{\Delta_1}{}{\Delta_2}{}{\Delta_{\delta_1}}{} \DrawVertex{\Delta_3}{}{\Delta_4}{}{\Delta_{\delta_1}}{}
\DrawEdge{\Delta_{\delta_1}}{} &\stackrel{!}{=} 1\cr 
 \DrawVertex{\Delta_1}{}{\Delta_2}{}{\Delta_{\delta_1}}{1} \DrawVertex{\Delta_3}{}{\Delta_4}{}{\Delta_{\delta_1}}{1}
\DrawEdge{\Delta_{\delta_1}}{1} &\stackrel{!}{=} { -\Delta_{\delta_11,2} \: \Delta_{\delta_12,1} \: \Delta_{\delta_13,4} \: \Delta_{\delta_14,3} \over \Delta_{\delta_1} (\Delta_{\delta_1}+1) (\Delta_{\delta_1}-{d\over 2}+1)  (\Delta_{\delta_1}-{d\over 2}+2)}\,.
}
We assign the simplest rules for the vertex and edge factors which reproduces~\eno{FourRqr}:\footnote{The overall minus signs in~\eno{FourEV} are included for consistency with higher-point examples which we will turn to next.}
\eqn{FourEV}{
\DrawVertex{\Delta_1}{}{\Delta_2}{}{\Delta_{\delta_1}}{} = 1 \qquad \qquad &
\DrawVertex{\Delta_1}{}{\Delta_2}{}{\Delta_{\delta_1}}{1} = -\Delta_{\delta_11,2} \: \Delta_{\delta_12,1} \cr 
\DrawEdge{\Delta_{\delta_1}}{} = 1 \qquad \qquad  &
\DrawEdge{\Delta_{\delta_1}}{1} = {-1 \over  \Delta_{\delta_1} (\Delta_{\delta_1}+1) (\Delta_{\delta_1}-{d\over 2}+1)  (\Delta_{\delta_1}-{d\over 2}+2) }\,.
}

Moving on to the five-point relation~\eno{Rych5}, we begin by determining the coefficients that appear in the dimensional reduction.
We first substitute~\eno{bareblockDef} into~\eno{ToFind} and do the change of variables as described above to obtain an infinite set of linear equations with constant coefficients for the  $\lambda_{\ldots}^{(t)}$ coefficients at each order in powers of the cross-ratios.  Solving the first few linear equations (e.g.\ those obtained at the first few orders in powers of $u_1^{k_1} u_2^{k_2}$) we obtain
\eqn{FiveRqr}{
 \figCFive{\Delta_1}{\Delta_2}{\Delta_3}{\Delta_4}{\Delta_5}{\Delta_{\delta_1}}{}{\Delta_{\delta_2}}{} &=1 \cr 
 \figCFive{\Delta_1}{\Delta_2}{\Delta_3}{\Delta_4}{\Delta_5}{\Delta_{\delta_1}}{1}{\Delta_{\delta_2}}{} &=  \figCFour{\Delta_1}{\Delta_2}{\Delta_3}{\Delta_{\delta_2}}{\Delta_{\delta_1}}{1} \cr 
 \figCFive{\Delta_1}{\Delta_2}{\Delta_3}{\Delta_4}{\Delta_5}{\Delta_{\delta_1}}{}{\Delta_{\delta_2}}{1} &=  \figCFour{\Delta_{\delta_1}}{\Delta_3}{\Delta_4}{\Delta_5}{\Delta_{\delta_2}}{1}\,,
}
and
\eqn{FiveRqr2}{
&  \figCFive{\Delta_1}{\Delta_2}{\Delta_3}{\Delta_4}{\Delta_5}{\Delta_{\delta_1}}{1}{\Delta_{\delta_2}}{1} \cr 
&=  { -\Delta_{\delta_11,2}  \Delta_{\delta_12,1} \Delta_{\delta_1 \delta_2,3} (\Delta_{\delta_1\delta_2,3}+1) (\Delta_{\delta_1\delta_23,} -h+1) (\Delta_{\delta_1\delta_2,3} -h+2)  \Delta_{\delta_2 4,5} \Delta_{\delta_2 5,4} \over \Delta_{\delta_1} (\Delta_{\delta_1}+1)  ( \Delta_{\delta_1}-h+1) ( \Delta_{\delta_1}-h+2)  \Delta_{\delta_2} (\Delta_{\delta_2}+1) (\Delta_{\delta_2}-h+1) (\Delta_{\delta_2}-h+2)} \,.
}
As a non-trivial check of the ansatz above, the solution above is indeed independent of the dummy parameters $k_i, j_{rs}$. This observation justifies why we could solve~\eno{ToFind} order by order in powers of the cross-ratios.
In fact while these coefficients were obtained by solving linear equations extracted from the first few orders in powers of cross-ratios $u_1^{k_1} u_2^{k_2}$ (say for values of $k_1, k_2$ up to 3), they satisfy~\eno{ToFind} to arbitrary orders in powers of cross-ratios.\footnote{We checked this statement analytically by verifying that~\eno{ToFind} is solved by~\eno{FiveRqr}-\eno{FiveRqr2} order by order in power of cross-ratios $u_1^{k_1} u_2^{k_2}$ to very high values of $\{k_1, k_2\}$.}

Interestingly, the new five-point coefficients in~\eno{FiveRqr} are given {\it solely} in terms of vertex and edge factors already determined in~\eno{FourEV}. 
It is satisfying to see that the simple assignment of vertex and edge factors chosen in~\eno{FourEV} manifestly reproduces the five-point coefficients in~\eno{FiveRqr}.
On the other hand, factorizing the coefficient on the left hand side of~\eno{FiveRqr2} using the ansatz above, 
\eqn{}{
\figCFive{\Delta_1}{\Delta_2}{\Delta_3}{\Delta_4}{\Delta_5}{\Delta_{\delta_1}}{1}{\Delta_{\delta_2}}{1} =  \DrawVertex{\Delta_1}{}{\Delta_2}{}{\Delta_{\delta_1}}{1} 
\DrawVertex{\Delta_{\delta_1}}{1}{\Delta_{\delta_2}}{1}{\Delta_3}{}
 \DrawVertex{\Delta_4}{}{\Delta_5}{}{\Delta_{\delta_2}}{1}
 \DrawEdge{\Delta_{\delta_1}}{1}  \DrawEdge{\Delta_{\delta_2}}{1} 
}
and substituting in~\eno{FourEV} we can solve for the undetermined vertex factor with two dashed-red edges attached, to obtain
\eqn{FiveEV}{
\DrawVertex{\Delta_{\delta_1}}{1}{\Delta_{\delta_2}}{1}{\Delta_3}{} =  -\Delta_{\delta_1\delta_2,3}\: (\Delta_{\delta_1\delta_2,3}+1) \left(\Delta_{\delta_1\delta_23,}-h+1\right)  \left(\Delta_{\delta_1\delta_2,3}-h+2\right) .
}

Finally, we can repeat the same exercise for the six-point block in the OPE/snowflake channel, which satisfies~\eno{Rych6OPE}. 
Note that all but one of the eight coefficients on the right hand side of~\eno{Rych6OPE} are fully determined in terms of previously fixed edge and vertex factors. We can check that the solution obtained from solving the linear equations is consistent with these constituent factors. Thus the exercise of determining seven of the coefficients serves as a consistency check of the ansatz, as well as of the previously obtained edge and vertex factors.
Upon factorization, the final coefficient includes the undetermined vertex with three dashed-red edges attached to it, which we can solve for using previously determined vertex and edge factors to obtain
\eqn{SixEV}{
\DrawVertex{\Delta_{\delta_1}}{1}{\Delta_{\delta_2}}{1}{\Delta_{\delta_3}}{1} = \Delta_{\delta_1\delta_2,\delta_3}\: \Delta_{\delta_2\delta_3,\delta_1} \: \Delta_{\delta_3\delta_1,\delta_2} \left(\Delta_{\delta_1\delta_2\delta_3,} - h +1\right) \left(\Delta_{\delta_1\delta_2\delta_3,} -h +2\right) \left(\Delta_{\delta_1\delta_2\delta_3,} - 2h +3\right).
}
It is trivial to check that these factors solve~\eno{ToFind} to any order in powers of cross-ratios.\footnote{In practice, we checked this to very large values of the integral parameters $k_1, k_2$ and $k_3$ which appear as exponents of the cross-ratios $u_1^{k_1} u_2^{k_2} u_3^{k_3}$ in~\eno{bareblockDef}.}

\vspace{0.75em}
Let's take stock of the calculations and results above: We determined all the edge and vertex factors, presented earlier in~\eno{VertexRules}-\eno{EdgeRules}, which collectively form the building blocks of the coefficients in the dimensional reduction relation~\eno{ToFind} for any $n$-point conformal block in an arbitrary topology. 
These factors were determined by solving linear equations obtained from~\eno{ToFind} order by order in powers of cross-ratios, and we needed to consider only three explicit cases to fix all factors: the four-point block, the five-point block and the six-point block in the OPE/snowflake topology.\footnote{Explicitly, we used the Feynman rules to substitute in the blocks in~\eno{ToFind}.}
Particularly, it was sufficient to only consider a finite set of linear equations to fix the factors; the full, infinite set of linear equations corresponding to arbitrary powers of cross-ratios in~\eno{ToFind} furnished a system of over-constrained linear equations, providing infinitely many consistency checks of the solution.
In practice we tested the solutions to a high but finite order.

Having determined the rules for the full set of building blocks, all other higher-point blocks (for any choice of topology) are guaranteed to satisfy dimensional reduction relation~\eno{ToFind} with coefficients given by the previously determined rules, since we have exhausted the full set of building blocks. 
Without the benefit of hindsight afforded by the proof above, this is a surprising consequence. 
It also serves as a highly non-trivial consistency requirement since no adjustable vertex or edge factors remain to ensure that all higher-point blocks satisfy the proposed dimensional reduction relations.
As a non-trivial consistency check, we can test whether the rules for the coefficients correctly reproduce the coefficients in~\eno{ToFind} for other toplogies not used in determining the rules: E.g.\ the six-point block in the comb topology, the two seven-point topologies, and all four eight-point topologies. 
We find that they do. 
In appendix~\ref{APP:EXAMPLES} we write down some of these explicit dimensional reduction relations and show all the topologies we explicitly tested.

\subsection{OPE and unit operator limits}
\label{OPEUNIT}

Finally, we close this section by performing two more consistency checks, namely we show that the dimensional reduction relation is consistent with the OPE limit and the unit operator limit.
As we'll see, the graphical  Feynman-like structure of the coefficients in the dimensional reduction relation simplifies the proofs considerably.

For the OPE limit, it is helpful to start with a concrete example for intuition. 
Consider the five-point dimensional reduction~\eno{Rych5}, more precisely, the conformal block version of it,
\eqn{Rych5W}{
W^{(d-2)}_{\Delta_{\delta_1},\Delta_{\delta_2}} &= \figCFive{\Delta_1}{\Delta_2}{\Delta_3}{\Delta_4}{\Delta_5}{\Delta_{\delta_1}}{}{\Delta_{\delta_2}}{} W^{(d)}_{\Delta_{\delta_1},\Delta_{\delta_2}} 
+ \figCFive{\Delta_1}{\Delta_2}{\Delta_3}{\Delta_4}{\Delta_5}{\Delta_{\delta_1}}{1}{\Delta_{\delta_2}}{} W^{(d)}_{\Delta_{\delta_1}+2,\Delta_{\delta_2}} \cr 
&+ \figCFive{\Delta_1}{\Delta_2}{\Delta_3}{\Delta_4}{\Delta_5}{\Delta_{\delta_1}}{}{\Delta_{\delta_2}}{1} W^{(d)}_{\Delta_{\delta_1},\Delta_{\delta_2}+2} 
+ \figCFive{\Delta_1}{\Delta_2}{\Delta_3}{\Delta_4}{\Delta_5}{\Delta_{\delta_1}}{1}{\Delta_{\delta_2}}{1} W^{(d)}_{\Delta_{\delta_1}+2,\Delta_{\delta_2}+2} \,,
}
and take the OPE limit $x_2 \to x_1$ on both sides.
The left hand side is simply given by 
\eqn{}{
LHS \stackrel{x_2 \to x_1}{\to} (x_{12}^2)^{\Delta_{\delta_1,12}}\: W_{\Delta_{\delta_2}}^{(d-2)}
}
where the external dimensions of the resulting four-point conformal block $W_{\Delta_{\delta_2}}^{(d)}$ are $\Delta_{\delta_1}, \Delta_3$ on one side of the exchanged operator $\Delta_{\delta_2}$ and $\Delta_4, \Delta_5$ on the other side.
On the right hand side of~\eno{Rych5}, the OPE limit results in
\eqn{}{
RHS &\stackrel{x_2 \to x_1}{\to} \figCFive{\Delta_1}{\Delta_2}{\Delta_3}{\Delta_4}{\Delta_5}{\Delta_{\delta_1}}{}{\Delta_{\delta_2}}{} 
(x_{12}^2)^{\Delta_{\delta_1,12}}\: W^{(d)}_{\Delta_{\delta_2}} 
+ \figCFive{\Delta_1}{\Delta_2}{\Delta_3}{\Delta_4}{\Delta_5}{\Delta_{\delta_1}}{1}{\Delta_{\delta_2}}{}
(x_{12}^2)^{\Delta_{\delta_1,12}+1}\: W^{(d)}_{\Delta_{\delta_2}} \cr 
&+ \figCFive{\Delta_1}{\Delta_2}{\Delta_3}{\Delta_4}{\Delta_5}{\Delta_{\delta_1}}{}{\Delta_{\delta_2}}{1}
(x_{12}^2)^{\Delta_{\delta_1,12}}\: W^{(d)}_{\Delta_{\delta_2}+2} 
+ \figCFive{\Delta_1}{\Delta_2}{\Delta_3}{\Delta_4}{\Delta_5}{\Delta_{\delta_1}}{1}{\Delta_{\delta_2}}{1}
(x_{12}^2)^{\Delta_{\delta_1,12}+1}\: W^{(d)}_{\Delta_{\delta_2}+2}\,.
}
We see that in the limit $x_2 \to x_1$, the second and fourth terms on the right hand side  contain an extra overall factor of $x_{12}^2$, and thus are sub-leading in the vanishing limit and can be dropped.
Then equating the left and right hand sides, we get
\eqn{FiveOPEeg}{
W_{\Delta_{\delta_2}}^{(d-2)} &= \figCFive{\Delta_1}{\Delta_2}{\Delta_3}{\Delta_4}{\Delta_5}{\Delta_{\delta_1}}{}{\Delta_{\delta_2}}{} 
 W^{(d)}_{\Delta_{\delta_2}} 
+ \figCFive{\Delta_1}{\Delta_2}{\Delta_3}{\Delta_4}{\Delta_5}{\Delta_{\delta_1}}{}{\Delta_{\delta_2}}{1}
 W^{(d)}_{\Delta_{\delta_2}+2} \cr 
 &= \figCFour{\Delta_{\delta_1}}{\Delta_3}{\Delta_4}{\Delta_5}{\Delta_{\delta_2}}{} 
 W^{(d)}_{\Delta_{\delta_2}} 
+ \figCFour{\Delta_{\delta_1}}{\Delta_3}{\Delta_4}{\Delta_5}{\Delta_{\delta_2}}{1}
 W^{(d)}_{\Delta_{\delta_2}+2}\,,
}
where in the second equality we used the Feynman-like factorization of the coefficients and the fact that a vertex factor with no dashed-red edges attached as well as a non-dashed-red internal edge factor simply contribute a factor of unity, and thus can be dropped.
For example, 
\eqn{}{
\figCFive{\Delta_1}{\Delta_2}{\Delta_3}{\Delta_4}{\Delta_5}{\Delta_{\delta_1}}{}{\Delta_{\delta_2}}{1} 
=  \DrawVertex{\Delta_1}{}{\Delta_2}{}{\Delta_{\delta_1}}{} \DrawEdge{\Delta_{\delta_1}}{} \figCFour{\Delta_{\delta_1}}{\Delta_3}{\Delta_4}{\Delta_5}{\Delta_{\delta_2}}{1} 
= \figCFour{\Delta_{\delta_1}}{\Delta_3}{\Delta_4}{\Delta_5}{\Delta_{\delta_2}}{1}.
}
The relation~\eno{FiveOPEeg} precisely reproduces  the four-point dimensional reduction~\eno{Rych4fig} (with appropriate relabeling).

More generally, the same steps can be applied to any pair of external legs incident at a common vertex in an $n$-point conformal block of arbitrary topology.
The blocks on both sides of the dimensional reduction reduce to $(n-1)$-point conformal blocks times overall factors of the form $(x_{ab}^2)^{\#}$ under the OPE limit $x_b \to x_a$.
In the dimensional reduction relation, half of the $2^{n-3}$ terms on the right hand side have the internal exchanged dimension incident at the vertex in question shifted up by two.
Thus under the OPE limit, such terms vanish due to an extra overall factor of $x_{ab}^2$ arising from the shifted dimension.
The remaining half of the terms are the ones where in the associated coefficient, the vertex under question has no incident dashed-red edges (i.e.\ the incident internal dimension is unshifted), thus the corresponding vertex and internal edge factors contribute a factor of unity and can be safely dropped to manifestly reproduce the correct graphical representations of the desired $(n-1)$-point coefficients.\footnote{The last step is what was done in the second equality of~\eno{FiveOPEeg}.}
This completes the check that the OPE limit acts consistently on the dimensional reduction. 

\vspace{.75em}
Now let's discuss the unit operator limit.
Once again it is useful to first consider the five-point dimensional reduction~\eno{Rych5}.
We will study two inequivalent unit operator limits: (a) $\Delta_1 = 0$ with $\Delta_{\delta_1} = \Delta_2$, and (b) $\Delta_3=0$ with $\Delta_{\delta_1} = \Delta_{\delta_2}$.

In case (a), the left hand side of~\eno{Rych5} becomes the four-point block $g^{(d-2)}_{\Delta_{\delta_2}}$ with external labels $\Delta_2, \Delta_3$ and $\Delta_4,\Delta_5$, 
\eqn{FiveUnitLHS}{
LHS = g^{(d-2)}_{\Delta_{\delta_2}} \,.
}
On the right hand side we get
\eqn{}{
RHS &= \figCFive{0}{\Delta_2}{\Delta_3}{\Delta_4}{\Delta_5}{\Delta_2}{}{\Delta_{\delta_2}}{} g^{(d)}_{\Delta_{\delta_2}} 
+ \figCFive{0}{\Delta_2}{\Delta_3}{\Delta_4}{\Delta_5}{\Delta_2}{1}{\Delta_{\delta_2}}{} \left[g^{(d)}_{\Delta_{\delta_1}+2,\Delta_{\delta_2}}\right]_{\substack{\Delta_1=0\\\Delta_{\delta_1}=\Delta_2}} \cr 
&+ \figCFive{0}{\Delta_2}{\Delta_3}{\Delta_4}{\Delta_5}{\Delta_2}{}{\Delta_{\delta_2}}{1} g^{(d)}_{\Delta_{\delta_2}+2} 
+ \figCFive{0}{\Delta_2}{\Delta_3}{\Delta_4}{\Delta_5}{\Delta_2}{1}{\Delta_{\delta_2}}{1} \left[g^{(d)}_{\Delta_{\delta_1}+2,\Delta_{\delta_2}+2}\right]_{\substack{\Delta_1=0\\\Delta_{\delta_1}=\Delta_2}}.
} 
The four-point blocks in the first and third terms above have external dimensions $\Delta_2,\Delta_3$ and $\Delta_4, \Delta_5$. 
Now we can use the Feynman factorization of the coefficients into edge and vertex factors to evaluate the coefficients.
Using the rules~\eno{VertexRules}-\eno{EdgeRules}, observe that 
\eqn{VertexSimplify}{
\DrawVertex{0}{}{\Delta_2}{}{\Delta_2}{1} = 0 \qquad  \qquad 
\DrawVertex{0}{}{\Delta_2}{}{\Delta_2}{} \DrawEdge{\Delta_2}{} = 1 \,,
}
from which we conclude that the second and fourth terms on the right hand side vanish, and the first and third terms simplify to give
\eqn{FiveUnitRHS}{
RHS = \figCFour{\Delta_2}{\Delta_3}{\Delta_4}{\Delta_5}{\Delta_{\delta_2}}{} g^{(d)}_{\Delta_{\delta_2}} 
+ \figCFour{\Delta_2}{\Delta_3}{\Delta_4}{\Delta_5}{\Delta_{\delta_2}}{1} g^{(d)}_{\Delta_{\delta_2}+2} \,.
}
Together,~\eno{FiveUnitLHS} and~\eno{FiveUnitRHS} recover the four-point dimensional reduction~\eno{Rych4fig}.

In case (b), setting $\Delta_3=0$ with $\Delta_{\delta_1}=\Delta_{\delta_2}$, the left hand side of~\eno{Rych5} reduces to the four-point block $g^{(d-2)}_{\Delta_{\delta_2}}$ with external dimensions $\Delta_1, \Delta_2$ and $\Delta_4, \Delta_5$,
\eqn{FiveUnitLHS2}{
LHS = g^{(d-2)}_{\Delta_{\delta_2}}\,.
}
On the right hand side, we get
\eqn{FiveUnitRHS2}{
RHS &= \figCFive{\Delta_1}{\Delta_2}{0}{\Delta_4}{\Delta_5}{\Delta_{\delta_2}}{}{\Delta_{\delta_2}}{} g^{(d)}_{\Delta_{\delta_2}} 
+ \figCFive{\Delta_1}{\Delta_2}{0}{\Delta_4}{\Delta_5}{\Delta_{\delta_2}}{1}{\Delta_{\delta_2}}{} \left[g^{(d)}_{\Delta_{\delta_1}+2,\Delta_{\delta_2}}\right]_{\substack{\Delta_3=0\\\Delta_{\delta_1}=\Delta_{\delta_2}}} \cr 
&+ \figCFive{\Delta_1}{\Delta_2}{0}{\Delta_4}{\Delta_5}{\Delta_{\delta_2}}{}{\Delta_{\delta_2}}{1} \left[g^{(d)}_{\Delta_{\delta_1},\Delta_{\delta_2}+2} \right]_{\substack{\Delta_3=0\\\Delta_{\delta_1}=\Delta_{\delta_2}}}
+ \figCFive{\Delta_1}{\Delta_2}{0}{\Delta_4}{\Delta_5}{\Delta_{\delta_2}}{1}{\Delta_{\delta_2}}{1} g^{(d)}_{\Delta_{\delta_2}+2} \,.
}
Once again using the Feynman factorization of the coefficients and the identities~\eno{VertexSimplify}, we notice that the second and third terms vanish, while the first term can be rewritten suggestively to give
\eqn{}{
RHS &= \figCFour{\Delta_1}{\Delta_2}{\Delta_4}{\Delta_5}{\Delta_{\delta_2}}{} g^{(d)}_{\Delta_{\delta_2}} 
+ \figCFive{\Delta_1}{\Delta_2}{0}{\Delta_4}{\Delta_5}{\Delta_{\delta_2}}{1}{\Delta_{\delta_2}}{1} g^{(d)}_{\Delta_{\delta_2}+2} \,.
}
The second coefficient above can be simplified further by using~\eno{VertexRules}-\eno{EdgeRules} and noting that
\eqn{}{
\DrawVertex{\Delta_{\delta_2}}{1}{\Delta_{\delta_2}}{1}{0}{} \DrawEdge{\Delta_{\delta_2}}{1} \DrawEdge{\Delta_{\delta_2}}{1} &=  { -\Delta_{\delta_2}\: (\Delta_{\delta_2}+1) \left(\Delta_{\delta_2}-h+1\right)  \left(\Delta_{\delta_2}-h+2\right) \over  \left(\Delta_{\delta_2} (\Delta_{\delta_2}+1) (\Delta_{\delta_2}-h+1)  (\Delta_{\delta_2}-h+2) \right)^2 } \cr 
 & = \DrawEdge{\Delta_{\delta_2}}{1}\,.
}
Thus we can rewrite the right hand side as
\eqn{}{
RHS &= \figCFour{\Delta_1}{\Delta_2}{\Delta_4}{\Delta_5}{\Delta_{\delta_2}}{} g^{(d)}_{\Delta_{\delta_2}} 
+ \figCFour{\Delta_1}{\Delta_2}{\Delta_4}{\Delta_5}{\Delta_{\delta_2}}{1} g^{(d)}_{\Delta_{\delta_2}+2} \,,
}
which together with~\eno{FiveUnitLHS2} recovers the four-point dimensional reduction.

Returning to a general $n$-point block, the unit operator limit at any external leg is equivalent to removing that leg to get an $(n-1)$-point block.
At any vertex with incident edges of dimensions, say $\Delta_1, \Delta_2$ and $\Delta_3$ where $\Delta_1$ is external and at least one of $\Delta_2$ and $\Delta_3$ is internal, mathematically this limit corresponds to setting $\Delta_1=0$ and simultaneously setting $\Delta_2 = \Delta_3$.\footnote{We will not worry about coordinate dependence in this discussion since it goes through as expected trivially.}

\begin{figure}
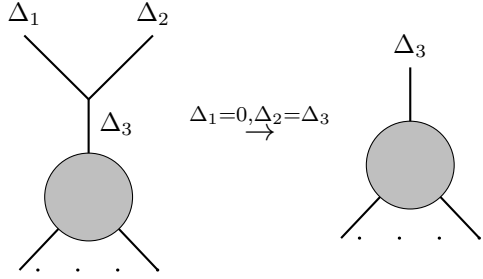

    \centering
    \[\musepic{\figNptUnitThree} \stackrel{\Delta_1=0, \Delta_2=\Delta_3}{\to} \musepic{\figNptUnitFour} \]
    \caption{The unit operator limit, $\Delta_1=0$ with $\Delta_2 = \Delta_3$ of an $n$-point block. The shaded blob admits an arbitrary topology. We have suppressed the position coordinates.}
    \label{fig:UnitOp2}
\end{figure}
Let's first consider the case when $\Delta_2$ is also external, so that $\Delta_3$ is an internal leg (see figure~\ref{fig:UnitOp2}).
Then taking the unit operator limit in the dimensional reduction formula, the left hand side will reduce trivially to an $(n-1)$-point bare conformal block in $(d-2)$ spacetime dimensions.
On the right hand side, half of the $2^{n-3}$ terms come with coefficients with the $\Delta_3$ leg shaded dashed-red, corresponding to a shifted exchanged dimension. 
In the Feynman representation of precisely these coefficients, there is always a factor of 
\eqn{VanishingVertex}{
\DrawVertex{\Delta_1}{}{\Delta_2}{}{\Delta_3}{1} \stackrel{\Delta_1=0, \Delta_2=\Delta_3}{=} 0 \,,
}
where we used the vertex factor from~\eno{VertexRules} to evaluate the limit.
Since the vertex factor vanishes, half of the terms in the dimensional reduction relation vanish.
It is now trivial to check that the other half of the terms are precisely the ones which arise in the dimensional reduction of the $(n-1)$-point obtained from taking the unit operator limit.\footnote{Just like in the discussion of the OPE limit, the vertex factor under question as well as the associated internal edge factor evaluate to unity and can be safely dropped. Doing this manifestly recovers the desired $(n-1)$-point coefficients.}

\begin{figure}
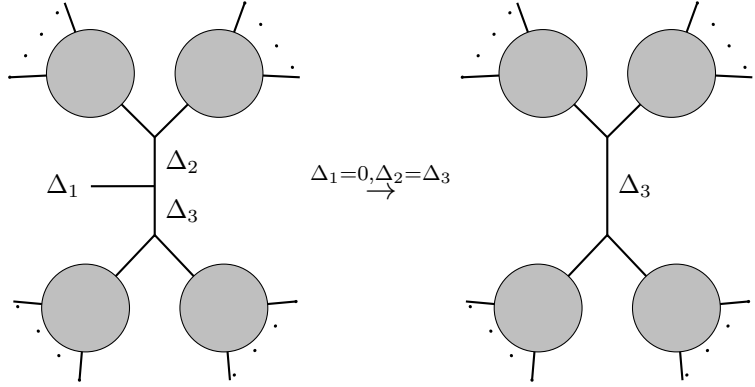

    \centering
    \[\musepic{\figNptUnitOne} \stackrel{\Delta_1=0, \Delta_2=\Delta_3}{\to} \musepic{\figNptUnitTwo} \]
    \caption{The unit operator limit, $\Delta_1=0$ with $\Delta_2 = \Delta_3$ of an $n$-point block. The shaded blobs admit arbitrary topologies.}
    \label{fig:UnitOp1}
\end{figure}

Now let's consider the other case when both $\Delta_2$ and $\Delta_3$ label internal legs and we take the unit operator limit $\Delta_1=0$ with $\Delta_2 = \Delta_3$ (see figure~\ref{fig:UnitOp1}).
Once again on the left hand side we get the appropriate $(n-1)$-point block in $(d-2)$ dimensions. 
We will now show that the right hand side also reduces to the appropriate linear combination of $d$-dimensional $(n-1)$-point blocks.
The $2^{n-3}$ terms on the right hand side of the $n$-point dimensional reduction can be partitioned into four sets of equal cardinality as shown in table~\ref{tb:Unit}.
\begin{table}[!h]
\begin{center}
\begin{tabular}{|c|c|c|c|c|}
    \hline
    shifted dimensions & $\emptyset$ & $\{ \Delta_2 \}$ & $\{\Delta_3\}$ & $\{\Delta_2, \Delta_3\}$ \\ \hline
    cardinality & $2^{n-5}$ & $2^{n-5}$ & $2^{n-5}$ & $2^{n-5}$ \\ \hline
\end{tabular}
\end{center}
\caption{Splitting the $2^{n-3}$ terms into four sets of equal cardinality.}
\label{tb:Unit}
\end{table}
In the table, $\emptyset$ is the set of terms where neither $\Delta_2$ nor $\Delta_3$ got shifted up by two, while the other three sets correspond to sets where one or both dimensions were shifted.

Each of the terms where $\Delta_2$ is shifted up (hence shaded dashed-red) but not $\Delta_3$, has a coefficient which by the Feynman rules contains a vertex (refer to figure~\ref{fig:UnitOp1})
\eqn{}{
\DrawVertex{\Delta_1}{}{\Delta_3}{}{\Delta_2}{1} \stackrel{\Delta_1=0, \Delta_2=\Delta_3}{=} 0 \,,
}
which vanishes for the same reason~\eno{VanishingVertex} did.
Likewise the terms where $\Delta_3$ is shifted up but not $\Delta_2$ also vanish. 
Thus one-half of the total $2^{n-3}$ terms vanish in the unit operator limit.

For terms where neither $\Delta_2$ nor $\Delta_3$ is shifted, each coefficient contains the following vertex and edge factors (refer to figure~\ref{fig:UnitOp1})
\eqn{}{
\DrawVertex{\Delta_2}{}{\Delta_3}{}{\Delta_1}{} \DrawEdge{\Delta_2}{} \DrawEdge{\Delta_3}{} \stackrel{\Delta_1=0, \Delta_2=\Delta_3}{=} 1 = \DrawEdge{\Delta_3}{}
}
where the second trivial equality emphasizes that in the unit operator limit, when we remove the external leg $\Delta_1$, in the resulting $(n-1)$-point block the original vertex disappears and we are left with a single exchanged operator (or internal leg) labeled $\Delta_3$ as shown in figure~\ref{fig:UnitOp1}.
Thus this set of $2^{n-5}$ terms exactly reproduces half of the terms appearing in the dimensional reduction of the reduced $(n-1)$-point block.

The remaining terms come from the final set of $n$-point terms where both $\Delta_2$ and $\Delta_3$ have been shifted. Here, each coefficient contains the following factors at the vertex where the unit operator limit is taken:
\eqn{}{
\DrawVertex{\Delta_2}{1}{\Delta_3}{1}{\Delta_1}{} \DrawEdge{\Delta_2}{1} \DrawEdge{\Delta_3}{1} &\stackrel{\Delta_1=0, \Delta_2=\Delta_3}{=}  { -\Delta_{3}\: (\Delta_{3}+1) \left(\Delta_{3}-{d\over 2}+1\right)  \left(\Delta_{3}-{d\over 2}+2\right) \over  \left(\Delta_{3} (\Delta_{3}+1) (\Delta_{3}-{d\over 2}+1)  (\Delta_{3}-{d\over 2}+2) \right)^2 } \cr 
 &\qquad = \DrawEdge{\Delta_3}{1}
}
where we used~\eno{VertexRules}-\eno{EdgeRules} to substitute in the vertex and edge factors.
The second equality and Feynman representation of the coefficients guarantees that these terms reproduce the remaining terms of the $(n-1)$-point dimensional reduction.

This concludes our proof of the fact that OPE limit and the unit operator limit act consistently on the higher-point dimensional reduction relations.

\section{Discussion}
\label{DISCUSS}

The close relation between the Parisi-Sourlas SCFT$_d$ and a dimensionally reduced CFT$_{d-2}$ led to the discovery of a remarkable finite series relation between ordinary $d$-dimensional and $(d-2)$-dimensional four-point conformal blocks~\cite{Kaviraj:2019tbg}.
In this paper we extended this relation to all scalar arbitrary-point blocks restricted to scalar exchanges.
This relation is captured in~\eno{ToFind} and summarized in the boxed text below~\eno{Rych5}.
One of the novel results of this paper is that the constant coefficients which appear as weights in the dimensional reduction relation for any conformal block are factorizable and completely specified by a compact set of Feynman-like rules~\eno{VertexRules}-\eno{EdgeRules}.
The derivation of these rules was greatly facilitated by recent work on determining an explicit power-series expansion for arbitrary scalar conformal blocks with the help of a set of Feynman rules of their own~\cite{Hoback:2020pgj}. 
In fact, in order to determine the rules obeyed by the constant coefficients of dimensional reduction relations for any higher-point block, we needed the knowledge of merely three basic conformal blocks: the four-point block, the five-point block and the six-point block in the OPE/snowflake channel.
These three topologies are the simplest conformal block topologies which together contain all three types of vertices in their unrooted binary tree representations: a vertex with exactly one incident internal edge, a vertex with exactly two incident internal edges, and a vertex with all three incident internal edges.
These vertices in turn give rise to Lauricella functions of one, two or three variables in the power-series expansion of conformal blocks, respectively~\cite{Hoback:2020pgj}. 
Consequently, vestiges of the Lauricella functions appear in the explicit factors in~\eno{VertexRules}.
While deriving the vertex and edge factors for the coefficients of dimensional reductions, we performed a variety of consistency checks at every step.
Notably, we showed that the OPE limit and unit operator limit act consistently on these dimensional relations, and explicitly verified that the vertex and edge factors -- completely fixed with the help of the three conformal blocks mentioned above, with no residual adjustable parameters -- do indeed correctly reproduce the dimensional reduction of all other $n$-point conformal blocks of all inequivalent typologies, at least for all $n \leq 8$.
Finally, in appendix~\ref{PARTIALWAVES} we also presented a simple, systematic method for determining the precise dimensional reduction relations obeyed by higher-point conformal partial waves.

This work suggests various interesting directions to further pursue.
It would be interesting to generalize the higher-point dimensional reduction relations to allow for spinning exchanges. 
The explicit form of the coefficients in the four-point dimensional reduction~\eno{Rych4Ell}~\cite{Kaviraj:2019tbg}, and the Feynman-like structure of the coefficients discovered in this paper for blocks with scalar exchanges strongly suggests that the coefficients in the more general spinning dimensional reduction relation will also satisfy nice factorization properties and Feynman-like rules.
In fact, just as for scalar exchanges, one should be able to fully determine the complete set of such rules simply by looking at four-point, five-point and six-point blocks in the OPE/snowflake channels, this time allowing for spinning exchanges.

One could ask whether it is possible to invert the dimensional reduction relations of this paper to find the higher-point analogs of the Dolan-Osborn dimensional uplift relations~\cite{Dolan:2003hv}. 
These relations give a finite series expansion of the $d$-dimensional conformal block in terms of $(d-2)$-dimensional conformal blocks (up to an overall cross-ratio dependent factor). 
Such relations are more practical from the point of view of finding closed-form expressions for conformal blocks, as they reduce the problem to finding closed-form expressions for conformal blocks in sufficiently low dimensions.

It would also be interesting to consider the implications of the higher-point dimensional reduction relations for higher-point Witten diagrams, along the lines recently discussed for the four-point case~\cite{Zhou:2020ptb}. 
In particular this might uncover new recursion relations and other interesting mathematical properties obeyed by higher-point bulk diagrams and perhaps make it possible to obtain closed-form expressions.

Finally, we would like to end with a comment on the relation of the present work to $p$-adic AdS/CFT~\cite{Gubser:2016guj,Heydeman:2016ldy}. 
In $p$-adic CFTs studied by Melzer~\cite{Melzer:1988he}, there are no descendants in the spectrum so that the conformal blocks are given simply by scaling blocks, i.e.\ the simplest possible power-laws~\cite{Gubser:2017tsi,Jepsen:2019svc}. 
From the point of view of the bare block in real CFTs, the $p$-adic bare block is obtained by collapsing all $k_i, j_{rs}$ sums in~\eno{bareblockDef} to the leading/zeroth order term.
In such a case, the block is independent of the spacetime dimension $d$, so that the dimensional reduction relations are trivial: Solely the coefficient in~\eno{ToFind} with all $m_i=0$ survives (and equals unity), while all other coefficients vanish.
This reaffirms that the non-trivial and interesting structure of dimensional reduction relations in conventional (real) CFTs studied in this paper is due in large parts to the presence and interplay of descendants in every conformal multiplet. 
It would be interesting to explore this point further to see whether there is a compelling physical explanation for the existence of finite term dimensional reductions which does not rely on Parisi-Sourlas supersymmetry.

 \subsection*{Acknowledgments}
 
 We thank J.-F.\ Fortin, W.-J.\ Ma and W.\ Skiba for useful discussions.

\appendix

\section{Dimensional reduction of conformal partial waves}
\label{PARTIALWAVES}

In this appendix we demonstrate how to obtain higher-point dimensional reduction relations for conformal {\it partial waves}. 
The $d$-dimensional four-point conformal partial associated with the conformal multiplet of dimension $\Delta_{\delta_1}$ and spin $\ell=0$ is defined by the following conformal integral~\cite{Ferrara:1972uq,Ferrara:1972ay,Ferrara:1972xe,Ferrara:1973vz,SimmonsDuffin:2012uy}
\eqn{4PartialWaveInt}{
\Psi_{\Delta_{\delta_1}}^{(d)[\Delta_1,\Delta_2,\Delta_3,\Delta_4]}(x_1, x_2, x_3, x_4) :=  \int_{y_1 \in \partial {\rm AdS}_{d+1}} \!\!\!\!\!\!\!\!\!\!\!\!\!\!\! \langle\langle {\cal O}_1(x_1) {\cal O}_2(x_2) {\cal O}_{\Delta_{\delta_1}}(y_1) \rangle\rangle \langle \langle \widetilde{ {\cal O}}_{\Delta_{\delta_1}}(y_1) {\cal O}_3(x_3) {\cal O}_4(x_4) \rangle \rangle\,,
}
where we use $\langle\langle \cdot \rangle\rangle$ to denote the purely kinematic part of the three-point function, i.e.\ devoid of the OPE coefficient,
and $\widetilde{ {\cal O}}_{\Delta}$ is the shadow operator of ${\cal O}_{\Delta}$ in $d$ dimensions,  defined by
\eqn{ShadowOp}{
\widetilde{\cal O}_\Delta(x) := \int_{y \in \partial{\rm AdS}_{d+1}} {1\over (x-y)^{2(d-\Delta)}} {\cal O}_\Delta(x)\,.
} 
It can be shown that the partial wave and conformal block are related via
\eqn{4cpw}{
\Psi_{\Delta_{\delta_1}}^{(d)[\Delta_1,\Delta_2,\Delta_3,\Delta_4]} =  K_{\widetilde{\Delta}_{\delta_1}}^{(d)\Delta_3,\Delta_4} W_0^{(4)} g_{\Delta_{\delta_1}}^{(d)}  + K_{\Delta_{\delta_1}}^{(d)\Delta_1,\Delta_2} W_0^{(4)} g_{\widetilde{\Delta}_{\delta_1}}^{(d)}\,,
}
where $\widetilde{\Delta}_{\delta_1} = d-\Delta_{\delta_1}$ is the shadow dimension, and 
\eqn{KcpwDef}{
K_{\widetilde{\Delta}_{\delta_1}}^{(d)\Delta_3,\Delta_4} = { \pi^{d/2} \Gamma({d\over 2}-\Delta_{\delta_1})  \Gamma(\Delta_{\delta_1 3,4}) \Gamma(\Delta_{\delta_1 4,3})  \over \Gamma(\Delta_{\delta_1}) \Gamma(\Delta_{\widetilde{\delta}_1 3,4}) \Gamma(\Delta_{\widetilde{\delta}_1 4,3}) } \,.
}

The four-point conformal partial wave satisfies a dimensional reduction schematically of the same form as the conformal block~\cite{Zhou:2020ptb}.
The precise coefficients differ from those appearing in~\eno{Rych4Ell} but are trivial to obtain starting from~\eno{Rych4Ell} and the relation between conformal blocks and partial waves~\eno{4cpw}.  
In this appendix we will not be interested in the coefficients but just the schematic structure. 
Schematically, we write the dimensional reduction relation (restricted to scalar exchanges) as
\eqn{4partialDimred}{
\Psi_{\Delta_{\delta_1}}^{(d-2)[\Delta_1,\Delta_2,\Delta_3,\Delta_4]} \sim \Psi_{\Delta_{\delta_1}}^{(d)[\Delta_1,\Delta_2,\Delta_3,\Delta_4]} +  \Psi_{\Delta_{\delta_1}+2}^{(d)[\Delta_1,\Delta_2,\Delta_3,\Delta_4]} \,.
}

Now recall that the five-point partial wave can be expressed as an integral over the four-point partial wave as follows:
\eqn{5PartialWaveInt}{
\Psi_{\Delta_{\delta_1}, \Delta_{\delta_2}}^{(d)[\Delta_1,\Delta_2,\Delta_3,\Delta_4,\Delta_5]}&(x_1, x_2, x_3, x_4, x_5) \cr 
:=&  \int_{y_2 \in \partial {\rm AdS}_{d+1}} \!\!\!\!\!\!\!\!\!\! \Psi^{(d)[\Delta_1,\Delta_2,\Delta_3,\Delta_{\delta_2}]}_{\Delta_{\delta_1}}(x_1, x_2, x_3, y_2)  \langle \langle \widetilde{ {\cal O}}_{\Delta_{\delta_2}}(y_2) {\cal O}_4(x_4) {\cal O}_5(x_5) \rangle \rangle\,.
}
We can consider the $(d-2)$-dimensional five-point partial wave and use~\eno{4partialDimred} to obtain,
\eqn{5PartialWaveInt1}{
\Psi_{\Delta_{\delta_1}, \Delta_{\delta_2}}^{(d-2)[\Delta_1,\Delta_2,\Delta_3,\Delta_4,\Delta_5]} &(x_1, \ldots, x_5) \cr 
=&  \int_{y_2 \in \partial {\rm AdS}_{d-1}} \!\!\!\!\!\!\!\!\!\! \Psi^{(d-2)[\Delta_1,\Delta_2,\Delta_3,\Delta_{\delta_2}]}_{\Delta_{\delta_1}}(x_1, x_2, x_3, y_2)  \langle \langle \hat{\widetilde{{\cal O}}}_{\Delta_{\delta_2}}(y_2) {\cal O}_4(x_4) {\cal O}_5(x_5) \rangle \rangle \cr 
 \sim& \int_{y_2 \in \partial {\rm AdS}_{d-1}} \!\!\!\!\!\!\!\!\!\! \Psi^{(d)[\Delta_1,\Delta_2,\Delta_3,\Delta_{\delta_2}]}_{\Delta_{\delta_1}}(x_1, x_2, x_3, y_2)  \langle \langle \hat{\widetilde{{\cal O}}}_{\Delta_{\delta_2}}(y_2) {\cal O}_4(x_4) {\cal O}_5(x_5) \rangle \rangle  \cr 
 +& \int_{y_2 \in \partial {\rm AdS}_{d-1}} \!\!\!\!\!\!\!\!\!\! \Psi^{(d)[\Delta_1,\Delta_2,\Delta_3,\Delta_{\delta_2}]}_{\Delta_{\delta_1}+2}(x_1, x_2, x_3, y_2)  \langle \langle \hat{\widetilde{{\cal O}}}_{\Delta_{\delta_2}}(y_2) {\cal O}_4(x_4) {\cal O}_5(x_5) \rangle \rangle  \,,
}
where $\hat{\widetilde{{\cal O}}}_{\Delta}$ denotes the shadow operator of ${\cal O}_{\Delta}$ in $(d-2)$ dimensions. 
Now substituting the integral representation~\eno{4PartialWaveInt} above, we recognize the $(d-2)$-dimensional $y_2$-integral over $\partial {\rm AdS}_{d-1}$ to be a $(d-2)$-dimensional four-point partial wave. 
Thus we get
\eqn{5PartialWaveInt2}{
&\Psi_{\Delta_{\delta_1}, \Delta_{\delta_2}}^{(d-2)[\Delta_1,\Delta_2,\Delta_3,\Delta_4,\Delta_5]}(x_1, \ldots, x_5)  \cr
 &\sim \int_{y_1 \in \partial {\rm AdS}_{d+1}}  \!\!\!\!\!\!\!\!\!\!\!\!\! \Psi^{(d-2)[d-\Delta_{\delta_1},\Delta_3,\Delta_4,\Delta_5]}_{\Delta_{\delta_2}}(y_1, x_3, x_4, x_5)  \langle \langle {\cal O}_1(x_1) {\cal O}_2(x_2) {\cal O}_{\Delta_{\delta_1}}(y_1)  \rangle \rangle  \cr 
 &+ \int_{y_1 \in \partial {\rm AdS}_{d+1}}  \!\!\!\!\!\!\!\!\!\!\!\!\! \Psi^{(d-2)[d-\Delta_{\delta_1}-2,\Delta_3,\Delta_4,\Delta_5]}_{\Delta_{\delta_2}}(y_1, x_3, x_4, x_5)  \langle \langle {\cal O}_1(x_1) {\cal O}_2(x_2)  { {\cal O}}_{\Delta_{\delta_1}+2}(y_1) \rangle \rangle  \,.
}
We can now use~\eno{4partialDimred} again to get a four-term relation
\eqn{5PartialWaveInt3}{
\Psi_{\Delta_{\delta_1}, \Delta_{\delta_2}}^{(d-2)[\Delta_1,\Delta_2,\Delta_3,\Delta_4,\Delta_5]}&(x_1, \ldots, x_5)  \cr 
 \sim& \int_{y_1 \in \partial {\rm AdS}_{d+1}}  \!\!\!\!\!\!\!\!\!\!\!\!\!\!\! \Psi^{(d)[d-\Delta_{\delta_1},\Delta_3,\Delta_4,\Delta_5]}_{\Delta_{\delta_2}}(y_1, x_3, x_4, x_5)  \langle \langle {\cal O}_1(x_1) {\cal O}_2(x_2) {\cal O}_{\Delta_{\delta_1}}(y_1)  \rangle \rangle  \cr 
 +& \int_{y_1 \in \partial {\rm AdS}_{d+1}}  \!\!\!\!\!\!\!\!\!\!\!\!\!\!\! \Psi^{(d)[d-\Delta_{\delta_1},\Delta_3,\Delta_4,\Delta_5]}_{\Delta_{\delta_2}+2}(y_1, x_3, x_4, x_5)  \langle \langle {\cal O}_1(x_1) {\cal O}_2(x_2) {\cal O}_{\Delta_{\delta_1}}(y_1)  \rangle \rangle  \cr 
 +& \int_{y_1 \in \partial {\rm AdS}_{d+1}}  \!\!\!\!\!\!\!\!\!\!\!\!\!\!\! \Psi^{(d)[d-\Delta_{\delta_1}-2,\Delta_3,\Delta_4,\Delta_5]}_{\Delta_{\delta_2}}(y_1, x_3, x_4, x_5)  \langle \langle {\cal O}_1(x_1) {\cal O}_2(x_2)  { {\cal O}}_{\Delta_{\delta_1}+2}(y_1) \rangle \rangle  \cr 
 +& \int_{y_1 \in \partial {\rm AdS}_{d+1}}  \!\!\!\!\!\!\!\!\!\!\!\!\!\!\! \Psi^{(d)[d-\Delta_{\delta_1}-2,\Delta_3,\Delta_4,\Delta_5]}_{\Delta_{\delta_2}+2}(y_1, x_3, x_4, x_5)  \langle \langle {\cal O}_1(x_1) {\cal O}_2(x_2)  { {\cal O}}_{\Delta_{\delta_1}+2}(y_1) \rangle \rangle \,.
}
Using~\eno{5PartialWaveInt} we recognize each term above as a $d$-dimensional five-point conformal partial wave, thus yielding the relation
\eqn{}{
\Psi_{\Delta_{\delta_1}, \Delta_{\delta_2}}^{(d-2)[\Delta_1,\Delta_2,\Delta_3,\Delta_4,\Delta_5]}&(x_1, \ldots, x_5)  \cr 
 \sim&\,  \Psi_{\Delta_{\delta_1}, \Delta_{\delta_2}}^{(d)[\Delta_1,\Delta_2,\Delta_3,\Delta_4,\Delta_5]}(x_1, \ldots, x_5) 
 +  \Psi_{\Delta_{\delta_1}, \Delta_{\delta_2}+2}^{(d)[\Delta_1,\Delta_2,\Delta_3,\Delta_4,\Delta_5]}(x_1, \ldots, x_5) \cr 
 +& \,  \Psi_{\Delta_{\delta_1}+2, \Delta_{\delta_2}}^{(d)[\Delta_1,\Delta_2,\Delta_3,\Delta_4,\Delta_5]}(x_1, \ldots, x_5) 
 +  \Psi_{\Delta_{\delta_1}+2, \Delta_{\delta_2}+2}^{(d)[\Delta_1,\Delta_2,\Delta_3,\Delta_4,\Delta_5]}(x_1, \ldots, x_5) \,,
}
which is precisely the partial wave analog of~\eno{Rych5}.

The derivation of the five-point partial wave relation makes it clear how to proceed systematically to obtain higher-point partial wave dimensional reduction relations.
One can repeat the previous calculations, this time starting with the integral representation of the six-point partial wave in terms of the five-point partial wave, and so on.
It should be clear that just like in the case of conformal blocks, there will be $2^{n-3}$ terms in the dimensional reduction, corresponding to all possible permutations of intermediate exchanged dimensions shifted up by two or not.\footnote{To obtain  higher-point dimensional reduction relations which include spinning exchanges using the iterative procedure of this appendix, one would first need to generalize dimensional reduction relations for four-point partial waves to allow at least one external spinning operator.}
One can also systematically obtain all the proportionality constants in the dimensional reduction relations by carefully tracking all factors in the calculation above.
Moreover, it is likely that there exist Feynman-like rules for determining the coefficients for arbitrary $n$-point partial waves of arbitrary topologies, although we do not pursue this direction here.

\section{Dimensional reduction of four-point blocks redux}
\label{HOGER}

 In ref.~\cite{Hogervorst:2016hal} a relation was derived for expressing $d$-dimensional four-point scalar conformal blocks in terms of an infinite series of blocks in a $(d-1)$-dimensional CFT. For simplicity, let's first focus on conformal blocks with identical external dimensions and scalar exchanges, in which case it was found that 
 \eqn{1dReduction}{
 g_{\Delta_\delta}^{(d)} = \sum_{n=0}^\infty {\cal A}^{(d)}_n(\Delta_\delta)\: g_{\Delta_\delta+2n}^{(d-1)}\,,
 }
 where the coefficients are fixed by conformal invariance and given by 
 \eqn{calADef}{
 {\cal A}^{(d)}_n(\Delta_\delta) = {(1/2)_n \over 4^n n!}  { (\Delta_\delta/2)_n^3 \over (\Delta_\delta - d/2 +1)_n (\Delta_\delta -d/2+1/2+n)_n  ((\Delta_\delta+1)/2)_n}\,.
 }

Applying the dimensional lift twice, one can express a $d$-dimensional block as a doubly-infinite series over $(d-2)$-dimensional blocks. However, after a simple shift in one of the dummy variables, one can isolate a cross-ratio independent sum, evaluating which one gets
 \eqn{2dReduction}{
 g_{\Delta_\delta}^{(d)} = \sum_{n=0}^\infty {\cal B}^{(d)}_n(\Delta_\delta)\: g_{\Delta_\delta+2n}^{(d-2)}\,,
 }
 where
 \eqn{calBDef}{
 {\cal B}^{(d)}_n(\Delta_\delta) = {1 \over 4^n}  { (\Delta_\delta/2)_n^3 \over (\Delta_\delta - d/2 +1)_{2n} ((\Delta_\delta+1)/2)_n} \,.
 }
Now observe that the ${\cal B}$ coefficients satisfy an extremely simple recursion relation:
\eqn{calBrecurse}{
{\cal B}_{n+1}^{(d)}(\Delta_\delta) = {\cal B}_1^{(d)}(\Delta_\delta) \: {\cal B}_n^{(d)}(\Delta_\delta+2)\,,
}
such that shifting $\Delta_\delta \to \Delta_{\delta}+2$ in~\eno{2dReduction} and multiplying by an overall factor, we  obtain
\eqn{}{
{\cal B}_1^{(d)}(\Delta_\delta)\: g_{\Delta_\delta+2}^{(d)} = \sum_{n=1}^\infty {\cal B}^{(d)}_n(\Delta_\delta)\: g_{\Delta_\delta+2n}^{(d-2)}\,.
}
Adding and subtracting $g^{(d-2)}_{\Delta_{\delta}}$ and recognizing the $d$-dimensional block using~\eno{2dReduction}, we get
\eqn{}{
{\cal B}_1^{(d)}(\Delta_\delta)\: g_{\Delta_\delta+2}^{(d)} = -g_{\Delta_\delta}^{(d-2)} + g_{\Delta_\delta}^{(d)}\,.
}
Thus the recursion relation~\eno{calBrecurse} enables miraculous cancellations and produces the (finite series) inverse of~\eno{2dReduction}, which is a special case of~\eno{4RychEll}-\eno{c20EllDef} with $\Delta_1=\Delta_2, \Delta_3=\Delta_4$ and $\ell=0$.

\vspace{.75em}

Now let's consider the case when the exchanged operator has spin $\ell \neq 0$. Then the $d-1\to d$ lift, for external dimensions $\Delta_1=\Delta_2, \Delta_3=\Delta_4$ is given by~\cite{Hogervorst:2016hal}
 \eqn{1dReductionEll}{
 g_{\Delta_\delta,\ell}^{(d)} = \sum_{n=0}^\infty \sum_j {\cal A}^{(d)}_{n,j}(\Delta_\delta,\ell)\: g_{\Delta_\delta+2n,j}^{(d-1)}\,,
 }
 where the $j$ sum runs over $j=\ell,\ell-2,\ldots, \ell-2 \mod 2$, and the series coefficients are known. Applying the lift twice, we get
  \eqn{}{
 g_{\Delta_\delta,\ell}^{(d)} = \sum_{n=0}^\infty \sum_j {\cal A}^{(d)}_{n,j}(\Delta_\delta,\ell)\sum_{m=0}^\infty \sum_k {\cal A}_{m,k}^{(d-1)}(\Delta_\delta+2n,j) g_{\Delta_\delta+2n+2m,k}^{(d-2)}\,,
 }
where the $k$ sum runs over $k=j,j-2,\ldots, j-2 \mod 2$. One can freely send $m\to m-n$, change the lower limit of the new $m$ sum to $0$ and switch the order of summation to obtain
  \eqn{}{
 g_{\Delta_\delta,\ell}^{(d)} = \sum_{m=0}^\infty \sum_{n=0}^\infty \sum_j \sum_k {\cal A}^{(d)}_{n,j}(\Delta_\delta,\ell)  {\cal A}_{m-n,k}^{(d-1)}(\Delta_\delta+2n,j) g_{\Delta_\delta+2m,k}^{(d-2)}\,.
 }
Now, rearranging the finite $j,k$ sums, we get
  \eqn{2dReductionEll}{
 g_{\Delta_\delta,\ell}^{(d)} &= \sum_{m=0}^\infty \sum_k \left(\sum_{n=0}^\infty \sum_j  {\cal A}^{(d)}_{n,j}(\Delta_\delta,\ell)  {\cal A}_{m-n,k}^{(d-1)}(\Delta_\delta+2n,j) \right) g_{\Delta_\delta+2m,k}^{(d-2)} \cr 
 &=: \sum_{m=0}^\infty \sum_k {\cal B}_{m,k}^{(d)}(\Delta_\delta,\ell) \: g_{\Delta_\delta+2m,k}^{(d-2)} \,,
 }
 where the $k$ sum is now over $k = \ell, \ell-2, \ldots, \ell-2 \mod 2$ and the $j$ sum now runs over $j= \ell, \ell-2, \ldots, k$.
 The series coefficients ${\cal B}_{m,k}$  are defined by the equation above, which allows us to write the $d-2 \to d$ dimension lift in terms of a single infinite series.
 We can compute these coefficients explicitly.
 Their exact form is tedious and unilluminating to include here.\footnote{The precise expression was found to be in terms of specific generalized hypergeometric functions ${}_8F_7, {}_9F_8$ and ${}_{10}F_{9}$. It is possible this can be simplified further, but we did not invest time on this exercise.}
 For special values of its parameters, it takes the form
 \eqn{}{
 {\cal B}_{m,\ell}^{(d)}(\Delta_\delta,\ell) &= {1 \over 4^m}  { (\Delta_\delta/2)_m  ((\Delta_\delta-1)/2)_m ((\Delta_{\delta} + \ell)/2)_m^2  \over (\Delta_\delta - d/2 +1)_{2m} ((\Delta_\delta+\ell+1)/2)_m ((\Delta_\delta+\ell-1)/2)_m } \cr 
 {\cal B}_{0,k}^{(d)}(\Delta_\delta,\ell) &=  {1 \over 2^{\ell-k}} { (-\ell)_{\ell-k} \over (2-d/2-\ell)_{\ell-k}} \,.
 }
 These coefficients satisfy recursion relations
 \eqn{calBsimpleRecur}{
  {\cal B}_{m+1,\ell}^{(d)}(\Delta_\delta,\ell) &=  {\cal B}_{1,\ell}^{(d)}(\Delta_\delta,\ell) \: {\cal B}_{m,\ell}^{(d)}(\Delta_\delta+2,\ell) \cr 
  {\cal B}_{0,k}^{(d)}(\Delta_\delta,\ell) &= {\cal B}_{0,\ell-2}^{(d)}(\Delta_\delta,\ell) \: {\cal B}_{0,k}^{(d)}(\Delta_\delta,\ell-2)\,.
 }
 
Just as in the $\ell=0$ case, we will consider variants of~\eno{2dReductionEll} with the dimension and spin of  the exchanged operator shifted variously.
 We can then add and subtract appropriate terms and make free use of~\eno{calBsimpleRecur} to obtain resummations which will help us invert~\eno{2dReductionEll} eventually.
For instance, one of the terms involved is,
\eqn{}{
{\cal B}_{1,\ell}^{(d)}(\Delta_\delta,\ell) g_{\Delta_\delta+2,\ell}^{(d)} &= {\cal B}_{1,\ell}^{(d)}(\Delta_\delta,\ell) \sum_{m=0}^\infty \sum_k {\cal B}_{m,k}^{(d)}(\Delta_\delta+2,\ell)g_{\Delta_\delta+2m+2,k}^{(d-2)}\cr 
 &= \sum_{m=1}^\infty  {\cal B}_{m,\ell}^{(d)}(\Delta_\delta,\ell)g_{\Delta_\delta+2m,\ell}^{(d-2)} + {\cal B}_{1,\ell}^{(d)}(\Delta_\delta,\ell) \sum_{m=0}^\infty \sum_{k=\ell-2,\ldots} {\cal B}_{m,k}^{(d)}(\Delta_\delta+2,\ell)g_{\Delta_\delta+2m+2,k}^{(d-2)} \cr 
 &= -g^{(d-2)}_{\Delta_\delta,\ell} + \sum_{m=0}^\infty  {\cal B}_{m,\ell}^{(d)}(\Delta_\delta,\ell)g_{\Delta_\delta+2m,\ell}^{(d-2)} \cr 
 & \quad + {\cal B}_{1,\ell}^{(d)}(\Delta_\delta,\ell) \sum_{m=0}^\infty \sum_{k=\ell-2,\ldots} {\cal B}_{m,k}^{(d)}(\Delta_\delta+2,\ell)g_{\Delta_\delta+2m+2,k}^{(d-2)} \,,
} 
where the sum with $k=\ell-2,\ldots$ denotes a shifted lower limit in the finite sum.
Likewise, one can consider other terms corresponding to the bare blocks with dimensions and spins shifted appropriately. Then using the recursion relations~\eno{calBsimpleRecur} and 
the following non-trivial recursion relation,
\eqn{calCDef}{
 {\cal B}_{m,k}^{(d)}(\Delta_\delta,\ell) &=  {\cal B}_{0,\ell-2}^{(d)}(\Delta_\delta,\ell) {\cal B}_{m,k}^{(d)}(\Delta_\delta,\ell-2) - {\cal B}_{0,\ell-2}^{(d)}(\Delta_\delta,\ell) {\cal B}_{1,\ell-2}^{(d)}(\Delta_\delta,\ell-2) {\cal B}_{m-1,k}^{(d)}(\Delta_\delta+2,\ell-2) \cr 
& + {\cal B}_{1,\ell}^{(d)}(\Delta_\delta,\ell) {\cal B}_{m-1,k}^{(d)}(\Delta_\delta+2,\ell) - {\cal B}_{1,\ell}^{(d)}(\Delta_\delta,\ell) {\cal B}_{0,\ell-2}^{(d)}(\Delta_\delta+2,\ell)  {\cal B}_{m-1,k}^{(d)}(\Delta_\delta+2,\ell-2) \cr 
&+  {\cal B}_{1,\ell-2}^{(d)}(\Delta_\delta,\ell) {\cal B}_{m-1,k}^{(d)}(\Delta_\delta+2,\ell-2) 
}
for $m=1,2,\ldots$ and $k=\ell-2,\ell-4,\ldots, \ell-2 \mod 2$, which can be checked numerically using the explicit form of the ${\cal B}_{m,k}^{(d)}$ coefficients, one obtains
\eqn{Rych4Hoger}{
& \left(  {\cal B}_{1,\ell-2}^{(d)}(\Delta_\delta,\ell)  - {\cal B}_{0,\ell-2}^{(d)}(\Delta_\delta,\ell) {\cal B}_{1,\ell-2}^{(d)}(\Delta_\delta,\ell-2) -{\cal B}_{1,\ell}^{(d)}(\Delta_\delta,\ell) {\cal B}_{0,\ell-2}^{(d)}(\Delta_\delta+2,\ell) \right) g_{\Delta_\delta+2,\ell-2}^{(d)} \cr 
 &+ {\cal B}_{1,\ell}^{(d)}(\Delta_\delta,\ell) g_{\Delta_\delta+2,\ell}^{(d)}  + {\cal B}_{0,\ell-2}^{(d)}(\Delta_\delta,\ell) g_{\Delta_\delta,\ell-2}^{(d)}  =  -g^{(d-2)}_{\Delta_\delta,\ell} + g_{\Delta_\delta,\ell}^{(d)}\,.
}
Equation~\eno{Rych4Hoger} is precisely the $d \to d-2$ dimensional reduction~\eno{4RychEll} of ref.~\cite{Kaviraj:2019tbg} for $\Delta_1 = \Delta_2, \Delta_3 = \Delta_4$. 
This provides a precise relation between the coefficients of the  infinite series expansion~\eno{2dReductionEll} and the coefficients of its finite inverse~\eno{4RychEll}:\footnote{When $\Delta_1 = \Delta_2, \Delta_3=\Delta_4$, the coefficient $c_{1,-1}$ in~\eno{4RychEll} vanishes, which is consistent with~\eno{Rych4Hoger} where the corresponding term did not appear.}
\eqn{}{
\centering
\begin{gathered}
c_{2,0} = -{\cal B}_{1,\ell}^{(d)}(\Delta_\delta,\ell)  
\qquad  \qquad 
c_{0,-2} = - {\cal B}_{0,\ell-2}^{(d)}(\Delta_\delta,\ell) 
\cr 
c_{2,-2} = -{\cal B}_{1,\ell-2}^{(d)}(\Delta_\delta,\ell)  + {\cal B}_{0,\ell-2}^{(d)}(\Delta_\delta,\ell) {\cal B}_{1,\ell-2}^{(d)}(\Delta_\delta,\ell-2) + {\cal B}_{1,\ell}^{(d)}(\Delta_\delta,\ell) {\cal B}_{0,\ell-2}^{(d)}(\Delta_\delta+2,\ell) \,,
\end{gathered}
}
which can be easily verified.

\section{More examples}
\label{APP:EXAMPLES}

In this appendix we write down the explicit dimensional reduction relations for some more higher-point examples. 
The coefficients are determined using the vertex and edge factors~\eno{VertexRules}-\eno{EdgeRules}.

\begin{figure}[!tb]
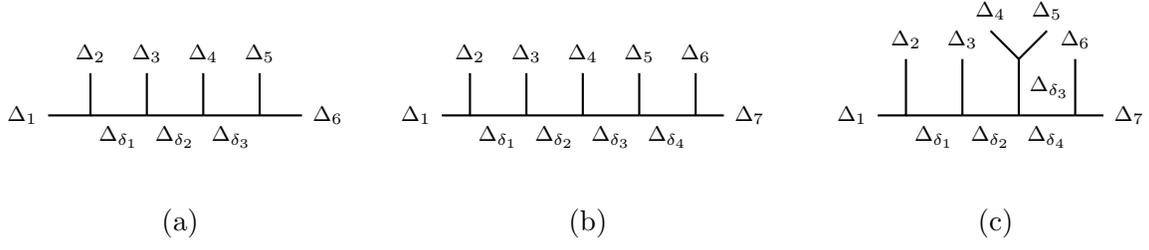

    \centering
    \begin{subfigure}[b]{0.32\textwidth}
    \[  \musepic{\figSixCombCB} \]
    \caption{}
    \label{fig:SixCombCB}
    \end{subfigure}
    \begin{subfigure}[b]{0.32\textwidth}
    \[  \musepic{\figSevenCombCB} \]
    \caption{}
    \label{fig:SevenCombCB}
    \end{subfigure}
    \begin{subfigure}[b]{0.32\textwidth}
     \[  \musepic{\figSevenMixedCB} \]
    \caption{}
    \label{fig:SevenMixedCB}
    \end{subfigure}
    \caption{(a) Six-point conformal block in the comb channel, and seven-point conformal blocks in the (b) comb channel and (c) ``mixed''/``extended snowflake'' channel. }
    \label{fig:SixSevenCB}
\end{figure}

The six-point block in the {\it comb} channel shown in figure~\ref{fig:SixCombCB}, for which an explicit power series representation in $d$ dimensions was first obtained in refs.~\cite{Parikh:2019dvm,Fortin:2019zkm}, satisfies:
\begingroup\makeatletter\def\f@size{10}\check@mathfonts 
\allowdisplaybreaks
\begin{align*}
& g^{(d-2)}_{\Delta_{\delta_1},\Delta_{\delta_2},\Delta_{\delta_3}} \\ 
&= \figCSixComb{\Delta_1}{\Delta_2}{\Delta_3}{\Delta_4}{\Delta_5}{\Delta_6}{\Delta_{\delta_1}}{}{\Delta_{\delta_2}}{}{\Delta_{\delta_3}}{} 
g^{(d)}_{\Delta_{\delta_1},\Delta_{\delta_2},\Delta_{\delta_3}} 
+ \figCSixComb{\Delta_1}{\Delta_2}{\Delta_3}{\Delta_4}{\Delta_5}{\Delta_6}{\Delta_{\delta_1}}{1}{\Delta_{\delta_2}}{}{\Delta_{\delta_3}}{} 
g^{(d)}_{\Delta_{\delta_1}+2,\Delta_{\delta_2},\Delta_{\delta_3}} \\ 
&+ \figCSixComb{\Delta_1}{\Delta_2}{\Delta_3}{\Delta_4}{\Delta_5}{\Delta_6}{\Delta_{\delta_1}}{}{\Delta_{\delta_2}}{1}{\Delta_{\delta_3}}{} 
g^{(d)}_{\Delta_{\delta_1},\Delta_{\delta_2}+2,\Delta_{\delta_3}}  
 + \figCSixComb{\Delta_1}{\Delta_2}{\Delta_3}{\Delta_4}{\Delta_5}{\Delta_6}{\Delta_{\delta_1}}{}{\Delta_{\delta_2}}{}{\Delta_{\delta_3}}{1} 
 g^{(d)}_{\Delta_{\delta_1},\Delta_{\delta_2},\Delta_{\delta_3}+2}  \\ 
& +\figCSixComb{\Delta_1}{\Delta_2}{\Delta_3}{\Delta_4}{\Delta_5}{\Delta_6}{\Delta_{\delta_1}}{1}{\Delta_{\delta_2}}{1}{\Delta_{\delta_3}}{}
g^{(d)}_{\Delta_{\delta_1}+2,\Delta_{\delta_2}+2,\Delta_{\delta_3}} 
+ \figCSixComb{\Delta_1}{\Delta_2}{\Delta_3}{\Delta_4}{\Delta_5}{\Delta_6}{\Delta_{\delta_1}}{}{\Delta_{\delta_2}}{1}{\Delta_{\delta_3}}{1}
g^{(d)}_{\Delta_{\delta_1},\Delta_{\delta_2}+2,\Delta_{\delta_3}+2} \\ 
& + \figCSixComb{\Delta_1}{\Delta_2}{\Delta_3}{\Delta_4}{\Delta_5}{\Delta_6}{\Delta_{\delta_1}}{1}{\Delta_{\delta_2}}{}{\Delta_{\delta_3}}{1} 
g^{(d)}_{\Delta_{\delta_1}+2,\Delta_{\delta_2},\Delta_{\delta_3}+2} 
+ \figCSixComb{\Delta_1}{\Delta_2}{\Delta_3}{\Delta_4}{\Delta_5}{\Delta_6}{\Delta_{\delta_1}}{1}{\Delta_{\delta_2}}{1}{\Delta_{\delta_3}}{1}
g^{(d)}_{\Delta_{\delta_1}+2,\Delta_{\delta_2}+2,\Delta_{\delta_3}+2}.
\stepcounter{equation}\tag{\theequation}
\label{Rych6comb}
\end{align*}
\endgroup

The seven-point block in the {\it mixed} channel depicted in figure~\ref{fig:SevenMixedCB}, for which an explicit power series representation in $d$ dimensions was first obtained in refs.~\cite{Fortin:2020bfq,Hoback:2020pgj}, satisfies a dimensional reduction with $2^{4}$ terms,
\begingroup\makeatletter\def\f@size{9}\check@mathfonts 
\allowdisplaybreaks
\begin{align*}
& g^{(d-2)}_{\Delta_{\delta_1},\Delta_{\delta_2},\Delta_{\delta_3},\Delta_{\delta_4}} \\ 
&= \figCSevenMix{\Delta_1}{\Delta_2}{\Delta_3}{\Delta_4}{\Delta_5}{\Delta_6}{\Delta_7}{\Delta_{\delta_1}}{}{\Delta_{\delta_2}}{}{\Delta_{\delta_3}}{}{\Delta_{\delta_4}}{}
g^{(d)}_{\Delta_{\delta_1},\Delta_{\delta_2},\Delta_{\delta_3},\Delta_{\delta_4}} 
+ \figCSevenMix{\Delta_1}{\Delta_2}{\Delta_3}{\Delta_4}{\Delta_5}{\Delta_6}{\Delta_7}{\Delta_{\delta_1}}{1}{\Delta_{\delta_2}}{}{\Delta_{\delta_3}}{}{\Delta_{\delta_4}}{}
g^{(d)}_{\Delta_{\delta_1}+2,\Delta_{\delta_2},\Delta_{\delta_3},\Delta_{\delta_4}} \\ 
&+ \figCSevenMix{\Delta_1}{\Delta_2}{\Delta_3}{\Delta_4}{\Delta_5}{\Delta_6}{\Delta_7}{\Delta_{\delta_1}}{}{\Delta_{\delta_2}}{1}{\Delta_{\delta_3}}{}{\Delta_{\delta_4}}{}
g^{(d)}_{\Delta_{\delta_1},\Delta_{\delta_2}+2,\Delta_{\delta_3},\Delta_{\delta_4}} 
+ \figCSevenMix{\Delta_1}{\Delta_2}{\Delta_3}{\Delta_4}{\Delta_5}{\Delta_6}{\Delta_7}{\Delta_{\delta_1}}{}{\Delta_{\delta_2}}{}{\Delta_{\delta_3}}{1}{\Delta_{\delta_4}}{}
g^{(d)}_{\Delta_{\delta_1},\Delta_{\delta_2},\Delta_{\delta_3}+2,\Delta_{\delta_4}} \\ 
&+ \figCSevenMix{\Delta_1}{\Delta_2}{\Delta_3}{\Delta_4}{\Delta_5}{\Delta_6}{\Delta_7}{\Delta_{\delta_1}}{}{\Delta_{\delta_2}}{}{\Delta_{\delta_3}}{}{\Delta_{\delta_4}}{1}
g^{(d)}_{\Delta_{\delta_1},\Delta_{\delta_2},\Delta_{\delta_3},\Delta_{\delta_4}+2} 
+ \figCSevenMix{\Delta_1}{\Delta_2}{\Delta_3}{\Delta_4}{\Delta_5}{\Delta_6}{\Delta_7}{\Delta_{\delta_1}}{1}{\Delta_{\delta_2}}{1}{\Delta_{\delta_3}}{}{\Delta_{\delta_4}}{}
g^{(d)}_{\Delta_{\delta_1}+2,\Delta_{\delta_2}+2,\Delta_{\delta_3},\Delta_{\delta_4}}  \\ 
&+ \figCSevenMix{\Delta_1}{\Delta_2}{\Delta_3}{\Delta_4}{\Delta_5}{\Delta_6}{\Delta_7}{\Delta_{\delta_1}}{1}{\Delta_{\delta_2}}{}{\Delta_{\delta_3}}{1}{\Delta_{\delta_4}}{}
g^{(d)}_{\Delta_{\delta_1}+2,\Delta_{\delta_2},\Delta_{\delta_3}+2,\Delta_{\delta_4}} 
+ \figCSevenMix{\Delta_1}{\Delta_2}{\Delta_3}{\Delta_4}{\Delta_5}{\Delta_6}{\Delta_7}{\Delta_{\delta_1}}{1}{\Delta_{\delta_2}}{}{\Delta_{\delta_3}}{}{\Delta_{\delta_4}}{1}
g^{(d)}_{\Delta_{\delta_1}+2,\Delta_{\delta_2},\Delta_{\delta_3},\Delta_{\delta_4}+2} \\ 
&+ \figCSevenMix{\Delta_1}{\Delta_2}{\Delta_3}{\Delta_4}{\Delta_5}{\Delta_6}{\Delta_7}{\Delta_{\delta_1}}{}{\Delta_{\delta_2}}{1}{\Delta_{\delta_3}}{1}{\Delta_{\delta_4}}{}
g^{(d)}_{\Delta_{\delta_1},\Delta_{\delta_2}+2,\Delta_{\delta_3}+2,\Delta_{\delta_4}} 
+\figCSevenMix{\Delta_1}{\Delta_2}{\Delta_3}{\Delta_4}{\Delta_5}{\Delta_6}{\Delta_7}{\Delta_{\delta_1}}{}{\Delta_{\delta_2}}{1}{\Delta_{\delta_3}}{}{\Delta_{\delta_4}}{1}
g^{(d)}_{\Delta_{\delta_1},\Delta_{\delta_2}+2,\Delta_{\delta_3},\Delta_{\delta_4}+2}  \\ 
&+ \figCSevenMix{\Delta_1}{\Delta_2}{\Delta_3}{\Delta_4}{\Delta_5}{\Delta_6}{\Delta_7}{\Delta_{\delta_1}}{}{\Delta_{\delta_2}}{}{\Delta_{\delta_3}}{1}{\Delta_{\delta_4}}{1}
g^{(d)}_{\Delta_{\delta_1},\Delta_{\delta_2},\Delta_{\delta_3}+2,\Delta_{\delta_4}+2} 
+ \figCSevenMix{\Delta_1}{\Delta_2}{\Delta_3}{\Delta_4}{\Delta_5}{\Delta_6}{\Delta_7}{\Delta_{\delta_1}}{1}{\Delta_{\delta_2}}{1}{\Delta_{\delta_3}}{1}{\Delta_{\delta_4}}{}
g^{(d)}_{\Delta_{\delta_1}+2,\Delta_{\delta_2}+2,\Delta_{\delta_3}+2,\Delta_{\delta_4}} \\ 
&+ \figCSevenMix{\Delta_1}{\Delta_2}{\Delta_3}{\Delta_4}{\Delta_5}{\Delta_6}{\Delta_7}{\Delta_{\delta_1}}{1}{\Delta_{\delta_2}}{1}{\Delta_{\delta_3}}{}{\Delta_{\delta_4}}{1}
g^{(d)}_{\Delta_{\delta_1}+2,\Delta_{\delta_2}+2,\Delta_{\delta_3},\Delta_{\delta_4}+2} 
+\figCSevenMix{\Delta_1}{\Delta_2}{\Delta_3}{\Delta_4}{\Delta_5}{\Delta_6}{\Delta_7}{\Delta_{\delta_1}}{1}{\Delta_{\delta_2}}{}{\Delta_{\delta_3}}{1}{\Delta_{\delta_4}}{1}
g^{(d)}_{\Delta_{\delta_1}+2,\Delta_{\delta_2},\Delta_{\delta_3}+2,\Delta_{\delta_4}+2} \\ 
&+\figCSevenMix{\Delta_1}{\Delta_2}{\Delta_3}{\Delta_4}{\Delta_5}{\Delta_6}{\Delta_7}{\Delta_{\delta_1}}{}{\Delta_{\delta_2}}{1}{\Delta_{\delta_3}}{1}{\Delta_{\delta_4}}{1}
g^{(d)}_{\Delta_{\delta_1},\Delta_{\delta_2}+2,\Delta_{\delta_3}+2,\Delta_{\delta_4}+2} 
+ \figCSevenMix{\Delta_1}{\Delta_2}{\Delta_3}{\Delta_4}{\Delta_5}{\Delta_6}{\Delta_7}{\Delta_{\delta_1}}{1}{\Delta_{\delta_2}}{1}{\Delta_{\delta_3}}{1}{\Delta_{\delta_4}}{1}
g^{(d)}_{\Delta_{\delta_1}+2,\Delta_{\delta_2}+2,\Delta_{\delta_3}+2,\Delta_{\delta_4}+2}\,. 
\stepcounter{equation}\tag{\theequation}
\label{Rych7mix}
\end{align*}
\endgroup

We also checked that the seven-point comb channel shown in~\ref{fig:SevenCombCB} and first obtained in refs.~\cite{Parikh:2019dvm,Fortin:2019zkm} satisfies its own dimensional reduction with $2^4$ terms. 
\begin{figure}[!tb]
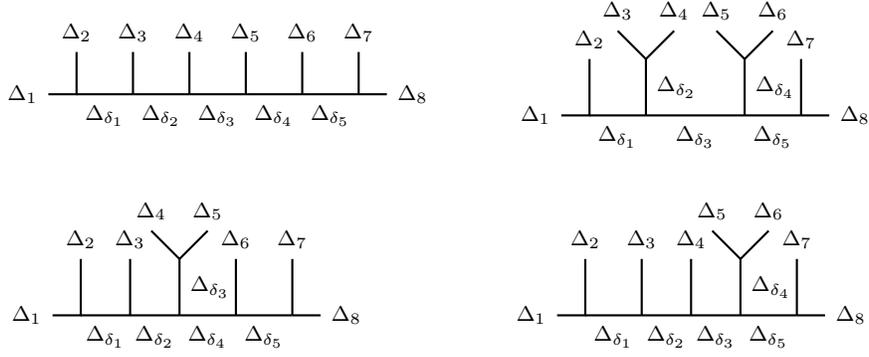

    \centering
    \[  \musepic{\figEightCombCB} \qquad \musepic{\figEightOPECB} \] 
    \[ \musepic{\figEightSymCB} \qquad\qquad \musepic{\figEightAsymCB} \]
    \caption{Eight-point conformal blocks in the four inequivalent topologies.}
    \label{fig:EightCB}
\end{figure}
Moreover, we explicitly checked that eight-point blocks of all four inequivalent topologies, shown in figure~\ref{fig:EightCB} and first written down in ref.~\cite{Hoback:2020pgj} with the help of Feynman rules, also satisfy the expected dimensional reductions with $2^5$ terms each.

\bibliographystyle{ssg}
\bibliography{main}

\begingroup\raggedright\begin{thebibliography}{10}

\bibitem{Ferrara:1973yt}
S.~Ferrara, A.~F. Grillo, and R.~Gatto, ``{Tensor representations of conformal
  algebra and conformally covariant operator product expansion},''
  \href{https://doi.org/10.1016/0003-4916(73)90446-6}{{\it Annals Phys.}} {\bf
  76} (1973) 161--188.

\bibitem{Polyakov:1974gs}
A.~M. Polyakov, ``{Nonhamiltonian approach to conformal quantum field
  theory},'' {\em Zh. Eksp. Teor. Fiz.} {\bf 66} (1974) 23--42.

\bibitem{Rattazzi:2008pe}
R.~Rattazzi, V.~S. Rychkov, E.~Tonni, and A.~Vichi, ``{Bounding scalar operator
  dimensions in 4D CFT},''
  \href{https://doi.org/10.1088/1126-6708/2008/12/031}{{\it JHEP}} {\bf 12}
  (2008) 031, \href{http://arxiv.org/abs/0807.0004}{{\tt 0807.0004}}.

\bibitem{Poland:2018epd}
D.~Poland, S.~Rychkov, and A.~Vichi, ``{The Conformal Bootstrap: Theory,
  Numerical Techniques, and Applications},''
  \href{https://doi.org/10.1103/RevModPhys.91.015002}{{\it Rev. Mod. Phys.}}
  {\bf 91} (2019), no.~1 15002, \href{http://arxiv.org/abs/1805.04405}{{\tt
  1805.04405}}.

\bibitem{Ferrara:1971vh}
S.~Ferrara, A.~F. Grillo, and R.~Gatto, ``{Manifestly conformal covariant
  operator-product expansion},'' \href{https://doi.org/10.1007/BF02770435}{{\it
  Lett. Nuovo Cim.}} {\bf 2S2} (1971) 1363--1369.

\bibitem{Ferrara:1972xe}
S.~Ferrara and G.~Parisi, ``{Conformal covariant correlation functions},''
  \href{https://doi.org/10.1016/0550-3213(72)90480-4}{{\it Nucl. Phys.}} {\bf
  B42} (1972) 281--290.

\bibitem{Ferrara:1973vz}
S.~Ferrara, A.~F. Grillo, G.~Parisi, and R.~Gatto, ``{Covariant expansion of
  the conformal four-point function},''
  \href{https://doi.org/10.1016/0550-3213(72)90587-1,
  10.1016/0550-3213(73)90467-7}{{\it Nucl. Phys.}} {\bf B49} (1972) 77--98.

\bibitem{Ferrara:1974ny}
S.~Ferrara, R.~Gatto, and A.~F. Grillo, ``{Properties of Partial Wave
  Amplitudes in Conformal Invariant Field Theories},''
  \href{https://doi.org/10.1007/BF02769009}{{\it Nuovo Cim.}} {\bf A26} (1975)
  226.

\bibitem{Dolan:2000ut}
F.~A. Dolan and H.~Osborn, ``{Conformal four point functions and the operator
  product expansion},''
  \href{https://doi.org/10.1016/S0550-3213(01)00013-X}{{\it Nucl. Phys.}} {\bf
  B599} (2001) 459--496, \href{http://arxiv.org/abs/hep-th/0011040}{{\tt
  hep-th/0011040}}.

\bibitem{Dolan:2003hv}
F.~A. Dolan and H.~Osborn, ``{Conformal partial waves and the operator product
  expansion},'' \href{https://doi.org/10.1016/j.nuclphysb.2003.11.016}{{\it
  Nucl. Phys.}} {\bf B678} (2004) 491--507,
  \href{http://arxiv.org/abs/hep-th/0309180}{{\tt hep-th/0309180}}.

\bibitem{Dolan:2011dv}
F.~A. Dolan and H.~Osborn, ``{Conformal Partial Waves: Further Mathematical
  Results},'' \href{http://arxiv.org/abs/1108.6194}{{\tt 1108.6194}}.

\bibitem{Penedones:2015aga}
J.~Penedones, E.~Trevisani, and M.~Yamazaki, ``{Recursion Relations for
  Conformal Blocks},'' \href{https://doi.org/10.1007/JHEP09(2016)070}{{\it
  JHEP}} {\bf 09} (2016) 070, \href{http://arxiv.org/abs/1509.00428}{{\tt
  1509.00428}}.

\bibitem{Kravchuk:2017dzd}
P.~Kravchuk, ``{Casimir recursion relations for general conformal blocks},''
  \href{https://doi.org/10.1007/JHEP02(2018)011}{{\it JHEP}} {\bf 02} (2018)
  011, \href{http://arxiv.org/abs/1709.05347}{{\tt 1709.05347}}.

\bibitem{Isachenkov:2016gim}
M.~Isachenkov and V.~Schomerus, ``{Superintegrability of $d$-dimensional
  Conformal Blocks},''
  \href{https://doi.org/10.1103/PhysRevLett.117.071602}{{\it Phys. Rev. Lett.}}
  {\bf 117} (2016), no.~7 071602, \href{http://arxiv.org/abs/1602.01858}{{\tt
  1602.01858}}.

\bibitem{Hijano:2015zsa}
E.~Hijano, P.~Kraus, E.~Perlmutter, and R.~Snively, ``{Witten Diagrams
  Revisited: The AdS Geometry of Conformal Blocks},''
  \href{https://doi.org/10.1007/JHEP01(2016)146}{{\it JHEP}} {\bf 01} (2016)
  146, \href{http://arxiv.org/abs/1508.00501}{{\tt 1508.00501}}.

\bibitem{Rosenhaus:2018zqn}
V.~Rosenhaus, ``{Multipoint Conformal Blocks in the Comb Channel},''
  \href{https://doi.org/10.1007/JHEP02(2019)142}{{\it JHEP}} {\bf 02} (2019)
  142, \href{http://arxiv.org/abs/1810.03244}{{\tt 1810.03244}}.

\bibitem{Meltzer:2019nbs}
D.~Meltzer, E.~Perlmutter, and A.~Sivaramakrishnan, ``{Unitarity Methods in
  AdS/CFT},'' \href{https://doi.org/10.1007/JHEP03(2020)061}{{\it JHEP}} {\bf
  03} (2020) 061, \href{http://arxiv.org/abs/1912.09521}{{\tt 1912.09521}}.

\bibitem{Fortin:2016lmf}
J.-F. Fortin and W.~Skiba, ``{Conformal Bootstrap in Embedding Space},''
  \href{https://doi.org/10.1103/PhysRevD.93.105047}{{\it Phys. Rev.}} {\bf D93}
  (2016), no.~10 105047, \href{http://arxiv.org/abs/1602.05794}{{\tt
  1602.05794}}.

\bibitem{Fortin:2019fvx}
J.-F. Fortin and W.~Skiba, ``{A recipe for conformal blocks},''
  \href{http://arxiv.org/abs/1905.00036}{{\tt 1905.00036}}.

\bibitem{Fortin:2019dnq}
J.-F. Fortin and W.~Skiba, ``{New methods for conformal correlation
  functions},'' \href{https://doi.org/10.1007/JHEP06(2020)028}{{\it JHEP}} {\bf
  06} (2020) 028, \href{http://arxiv.org/abs/1905.00434}{{\tt 1905.00434}}.

\bibitem{Fortin:2019zkm}
J.-F. Fortin, W.~Ma, and W.~Skiba, ``{Higher-Point Conformal Blocks in the Comb
  Channel},'' \href{https://doi.org/10.1007/JHEP07(2020)213}{{\it JHEP}} {\bf
  07} (2020) 213, \href{http://arxiv.org/abs/1911.11046}{{\tt 1911.11046}}.

\bibitem{Fortin:2020ncr}
J.-F. Fortin, W.-J. Ma, V.~Prilepina, and W.~Skiba, ``{Efficient Rules for All
  Conformal Blocks},'' \href{http://arxiv.org/abs/2002.09007}{{\tt
  2002.09007}}.

\bibitem{Fortin:2020yjz}
J.-F. Fortin, W.-J. Ma, and W.~Skiba, ``{Six-Point Conformal Blocks in the
  Snowflake Channel},'' \href{http://arxiv.org/abs/2004.02824}{{\tt
  2004.02824}}.

\bibitem{Fortin:2020bfq}
J.-F. Fortin, W.-J. Ma, and W.~Skiba, ``{Seven-Point Conformal Blocks in the
  Extended Snowflake Channel and Beyond},''
  \href{http://arxiv.org/abs/2006.13964}{{\tt 2006.13964}}.

\bibitem{Fortin:2020zxw}
J.-F. Fortin, W.-J. Ma, and W.~Skiba, ``{All Global One- and Two-Dimensional
  Higher-Point Conformal Blocks},'' \href{http://arxiv.org/abs/2009.07674}{{\tt
  2009.07674}}.

\bibitem{Goncalves:2019znr}
V.~Gonçalves, R.~Pereira, and X.~Zhou, ``{$20'$ Five-Point Function from
  $AdS_5\times S^5$ Supergravity},''
  \href{https://doi.org/10.1007/JHEP10(2019)247}{{\it JHEP}} {\bf 10} (2019)
  247, \href{http://arxiv.org/abs/1906.05305}{{\tt 1906.05305}}.

\bibitem{Parikh:2019ygo}
S.~Parikh, ``{Holographic dual of the five-point conformal block},''
  \href{https://doi.org/10.1007/JHEP05(2019)051}{{\it JHEP}} {\bf 05} (2019)
  051, \href{http://arxiv.org/abs/1901.01267}{{\tt 1901.01267}}.

\bibitem{Jepsen:2019svc}
C.~B. Jepsen and S.~Parikh, ``{Propagator identities, holographic conformal
  blocks, and higher-point AdS diagrams},''
  \href{https://doi.org/10.1007/JHEP10(2019)268}{{\it JHEP}} {\bf 10} (2019)
  268, \href{http://arxiv.org/abs/1906.08405}{{\tt 1906.08405}}.

\bibitem{Parikh:2019dvm}
S.~Parikh, ``{A multipoint conformal block chain in $d$ dimensions},''
  \href{https://doi.org/10.1007/JHEP05(2020)120}{{\it JHEP}} {\bf 05} (2020)
  120, \href{http://arxiv.org/abs/1911.09190}{{\tt 1911.09190}}.

\bibitem{Hoback:2020pgj}
S.~Hoback and S.~Parikh, ``{Towards Feynman rules for conformal blocks},''
  \href{http://arxiv.org/abs/2006.14736}{{\tt 2006.14736}}.

\bibitem{InPrep}
J.-F. Fortin, S.~Hoback, W.-J. Ma, S.~Parikh, and W.~Skiba {\em {in
  preparation}}.

\bibitem{Karateev:2017jgd}
D.~Karateev, P.~Kravchuk, and D.~Simmons-Duffin, ``{Weight Shifting Operators
  and Conformal Blocks},'' \href{https://doi.org/10.1007/JHEP02(2018)081}{{\it
  JHEP}} {\bf 02} (2018) 081, \href{http://arxiv.org/abs/1706.07813}{{\tt
  1706.07813}}.

\bibitem{Parisi:1979ka}
G.~Parisi and N.~Sourlas, ``{Random Magnetic Fields, Supersymmetry and Negative
  Dimensions},'' \href{https://doi.org/10.1103/PhysRevLett.43.744}{{\it Phys.
  Rev. Lett.}} {\bf 43} (1979) 744.

\bibitem{Kaviraj:2019tbg}
A.~Kaviraj, S.~Rychkov, and E.~Trevisani, ``{Random Field Ising Model and
  Parisi-Sourlas supersymmetry. Part I. Supersymmetric CFT},''
  \href{https://doi.org/10.1007/JHEP04(2020)090}{{\it JHEP}} {\bf 04} (2020)
  090, \href{http://arxiv.org/abs/1912.01617}{{\tt 1912.01617}}.

\bibitem{Zhou:2020ptb}
X.~Zhou, ``{How to Succeed at Witten Diagram Recursions without Really
  Trying},'' \href{http://arxiv.org/abs/2005.03031}{{\tt 2005.03031}}.

\bibitem{Hogervorst:2016hal}
M.~Hogervorst, ``{Dimensional Reduction for Conformal Blocks},''
  \href{https://doi.org/10.1007/JHEP09(2016)017}{{\it JHEP}} {\bf 09} (2016)
  017, \href{http://arxiv.org/abs/1604.08913}{{\tt 1604.08913}}.

\bibitem{oeis}
 The On-Line Encyclopedia of Integer Sequences, published electronically at
  \url{https://oeis.org/A000672}, sequence A000672.

\bibitem{Fitzpatrick:2011ia}
A.~L. Fitzpatrick, J.~Kaplan, J.~Penedones, S.~Raju, and B.~C. van Rees, ``{A
  Natural Language for AdS/CFT Correlators},''
  \href{https://doi.org/10.1007/JHEP11(2011)095}{{\it JHEP}} {\bf 11} (2011)
  095, \href{http://arxiv.org/abs/1107.1499}{{\tt 1107.1499}}.

\bibitem{Paulos:2011ie}
M.~F. Paulos, ``{Towards Feynman rules for Mellin amplitudes},''
  \href{https://doi.org/10.1007/JHEP10(2011)074}{{\it JHEP}} {\bf 10} (2011)
  074, \href{http://arxiv.org/abs/1107.1504}{{\tt 1107.1504}}.

\bibitem{Gubser:2016guj}
S.~S. Gubser, J.~Knaute, S.~Parikh, A.~Samberg, and P.~Witaszczyk, ``{$p$-adic
  AdS/CFT},'' \href{https://doi.org/10.1007/s00220-016-2813-6}{{\it Commun.
  Math. Phys.}} {\bf 352} (2017), no.~3 1019--1059,
  \href{http://arxiv.org/abs/1605.01061}{{\tt 1605.01061}}.

\bibitem{Heydeman:2016ldy}
M.~Heydeman, M.~Marcolli, I.~Saberi, and B.~Stoica, ``{Tensor networks,
  $p$-adic fields, and algebraic curves: arithmetic and the AdS$_3$/CFT$_2$
  correspondence},'' \href{https://doi.org/10.4310/ATMP.2018.v22.n1.a4}{{\it
  Adv. Theor. Math. Phys.}} {\bf 22} (2018) 93--176,
  \href{http://arxiv.org/abs/1605.07639}{{\tt 1605.07639}}.

\bibitem{Melzer:1988he}
E.~Melzer, ``{Nonarchimedean Conformal Field Theories},''
  \href{https://doi.org/10.1142/S0217751X89002065}{{\it Int. J. Mod. Phys.}}
  {\bf A4} (1989) 4877.

\bibitem{Gubser:2017tsi}
S.~S. Gubser and S.~Parikh, ``{Geodesic bulk diagrams on the Bruhat–Tits
  tree},'' \href{https://doi.org/10.1103/PhysRevD.96.066024}{{\it Phys. Rev.}}
  {\bf D96} (2017), no.~6 066024, \href{http://arxiv.org/abs/1704.01149}{{\tt
  1704.01149}}.

\bibitem{Ferrara:1972uq}
S.~Ferrara, A.~F. Grillo, G.~Parisi, and R.~Gatto, ``{The shadow operator
  formalism for conformal algebra. Vacuum expectation values and operator
  products},'' \href{https://doi.org/10.1007/BF02907130}{{\it Lett. Nuovo
  Cim.}} {\bf 4S2} (1972) 115--120.

\bibitem{Ferrara:1972ay}
S.~Ferrara, A.~F. Grillo, and G.~Parisi, ``{Nonequivalence between conformal
  covariant wilson expansion in euclidean and minkowski space},''
  \href{https://doi.org/10.1007/BF02815915}{{\it Lett. Nuovo Cim.}} {\bf 5S2}
  (1972) 147--151.

\bibitem{SimmonsDuffin:2012uy}
D.~Simmons-Duffin, ``{Projectors, Shadows, and Conformal Blocks},''
  \href{https://doi.org/10.1007/JHEP04(2014)146}{{\it JHEP}} {\bf 04} (2014)
  146, \href{http://arxiv.org/abs/1204.3894}{{\tt 1204.3894}}.

\end{thebibliography}\endgroup

\end{document}